\documentclass[11pt]{article}
 
\RequirePackage[a4paper]{geometry}
\usepackage[colorlinks,
            linkcolor=blue,
            anchorcolor=blue,
            citecolor=blue]{hyperref}
            
\usepackage{caption}
\usepackage{amsmath}
\usepackage{subfigure}
\usepackage{diagbox}
\usepackage{lscape,epsfig,graphicx,multirow}
\usepackage{indentfirst}
\usepackage[section]{placeins}
\usepackage{makecell}
\usepackage{color}
\usepackage{tabularx}
\usepackage[affil-it]{authblk}
{\normalsize {\large {\Large }}}
\title{\textbf{Reconstruction of Muon Bundle in the JUNO Central Detector}}
\author[1]{Cheng-Feng Yang}
\author[1]{Yong-Bo Huang\thanks{huangyb@gxu.edu.cn}}
\author[2]{Ji-Lei Xu\thanks{xujl@ihep.ac.cn}}
\author[2,3]{Di-Ru Wu}
\author[2]{Hao-Qi Lu}
\author[2]{Yong-Peng Zhang}
\author[2]{Wu-Ming Luo}
\author[2]{Miao He}
\author[1]{Guo-Ming Chen}
\author[1]{Si-Yuan Zhang}

\affil[1]{School of Physical Science and Technology, Guangxi University, Nanning 530004, China}
\affil[2]{Institute of High Energy Physics, Beijing 100049, China}
\affil[3]{University of Chinese Academy of Sciences, Beijing 100049, China}

\begin{document}
\captionsetup[figure]{labelfont={bf},labelformat={default},labelsep=space,name={Fig.}}
%Define new command for auto linefeed in table cell
%then, use &\tabincell{c}{}& to auto linefeed
\newcommand{\tabincell}[2]{
    \begin{tabular}{@{}#1@{}}#2\end{tabular}
    }
    
\maketitle

\begin{abstract}
	The Jiangmen Underground Neutrino Observatory (JUNO) is a multi-purpose neutrino experiment. One of the main goals is to determine the neutrino mass ordering by precisely measuring the energy spectrum of reactor antineutrinos.
	For reactor antineutrino detection, cosmogenic backgrounds such as $^9$Li/$^8$He and fast neutrons induced by cosmic muons should be rejected carefully by applying muon veto cuts, which requires good muon track reconstruction. With a 20~kton liquid scintillator detector, simulation shows the proportion of muon bundles to be around 8\% in the JUNO, while its reconstruction is rarely discussed in previous experiments. According to the charge response of the PMT array, this paper proposes an efficient algorithm for muon bundle track reconstruction. This is the first reconstruction of muon bundles in a large volume liquid scintillator detector. Additionally, the algorithm shows good performance and potential in reconstruction for both single muon and double muons. The spatial resolution of single muon reconstruction is 20~cm and the angular resolution is $0.5^\circ$. As for double muon reconstruction, the spatial resolution and angular resolution could be 30~cm and $1.0^\circ$, respectively. Moreover, this paper has also discussed muon classification and veto strategy.
\end{abstract}

\begin{flushleft}
    {\bf Keywords} JUNO, Liquid scintillator detector, Muon reconstruction, Muon bundle , Veto strategy
\end{flushleft}
	
\section{Introduction}

	The long-lived radioactive isotopes produced by energetic cosmic muons are one of the main backgrounds for large volume neutrino and dark matter detectors. Locating the detector in deep underground to provide an essential overburden is one effective approach to suppress muon-induced backgrounds. For a giant volume detector like JUNO, the muon rate is still much higher than the neutrino signal rate. Vetoing the entire volume for some time after a muon event is both uneconomical and would lose too much exposure. Since most cosmogenic isotopes are generated near the muon track, we adopt a strategy that only vetoes part of the volume near the (reconstructed) muon track. It has been proven that a cylindrical volume veto along the muon trajectory is a good approach~\cite{juno-yellowbook}.
	
\begin{figure}[!htb]
		\centering
		\includegraphics[width=0.8\textwidth]{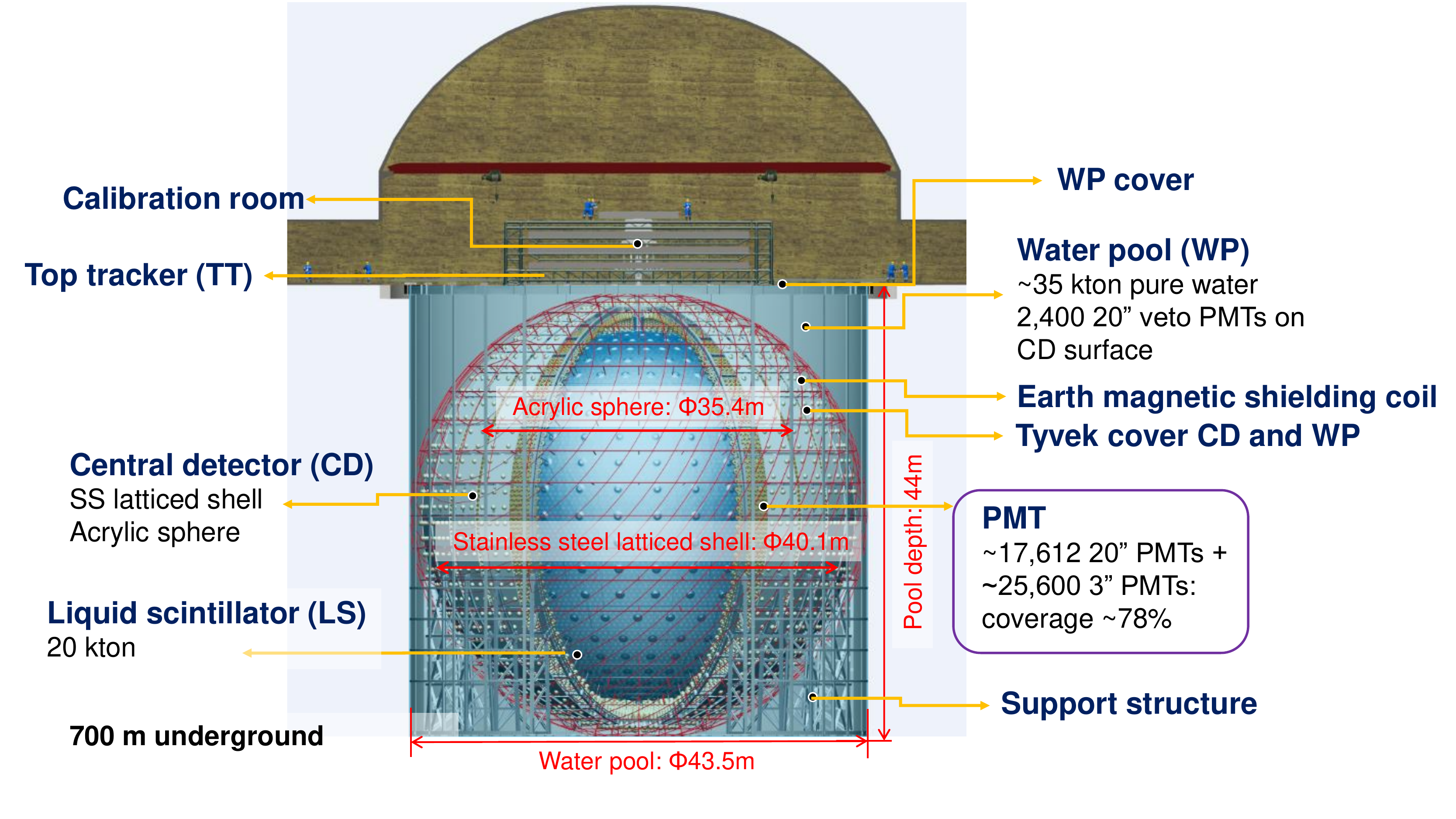}
		\caption{A schematic view of the JUNO detector.} 
		\label{fig:JUNOdetector} 
	\end{figure}

	The Jiangmen Underground Neutrino Observatory (JUNO)~\cite{JUNO-PPNP} is a multi-purpose neutrino experiment. One of the main goals is to determine the neutrino mass ordering by precisely measuring the energy spectrum of reactor antineutrinos at a site where the distance is ~53~km from the reactors of the Yangjiang and Taishan Nuclear Power Plants. Fig.~\ref{fig:JUNOdetector} is a schematic view of the JUNO detector. The Central Detector (CD) is a 20~kton liquid scintillator detector that uses 17,612 20-inch PMTs (LPMTs) and 25,600 3-inch PMTs (SPMTs) as photosensors. The water Cerenkov detector is a cylindrical water pool with 2400 20-inch PMTs and filled with 35~kton ultra-pure water as a muon veto and radioactive shielding. And the top tracker detector can be used to tag muon events. With powerful reactors at 53~km, the CD expects to detect about 83 Inverse Beta Decay (IBD) events caused by reactor per day, while the estimated cosmic muon rate is about %3.6 Hz (
	3.4 $\times$ 10$^5$ events per day. According to a preliminary antineutrino selection criterion~\cite{juno-yellowbook}, there are about 71 residual cosmogenic $^9$Li/$^8$He events per day, comparable to IBD events 73/day. Thus the $\beta-n$ decays from $^9$Li/$^8$He are the most serious correlated background to reactor antineutrinos and need to be further rejected by an efficient muon veto strategy. 
	
	There are several different kinds of muon tracks in JUNO spherical detector. According to the position relative to the detector and the behavior of the muon tracks, the muon tracks can be classified into the following categories, which are listed in table~\ref{tab-class}:
	
	(1) Through-going muon: corresponding to muon that goes through the CD and the distance between muon track and the center of the CD is less than 16~m; 
	
	(2) Clipping muon: corresponding to muon that goes through the CD but leaves only a short track in the detector (in other words, the distance between muon track and the center of CD is larger than 16~m); 
	
	(3) Stopping muon: corresponding to muon whose track stopped in CD;
	\begin{table*}
	\caption{The cluster number of different muon types}
	\label{tab-class}
		\renewcommand\arraystretch{1.6}%cm
		\centering
		\begin{tabular}{c|c|c|c}
			\hline
			\diagbox{Muon type}{Cluster Num.}{Muon behavior} & Through-going & Clipping & Stopping     \\ \hline
			Single muon & 2 clusters      & 1 cluster     & 1 cluster     \\ \hline
			Double muons & 4 clusters      & 2-3 clusters   & 2-3 clusters   \\ \hline
			Multiple muons & Over 4 clusters & Over 2 clusters & Over 2 clusters \\ \hline
		\end{tabular}
	\end{table*}
	
\begin{figure}[!htb]
		\centering
		\includegraphics[width=0.45\textwidth]{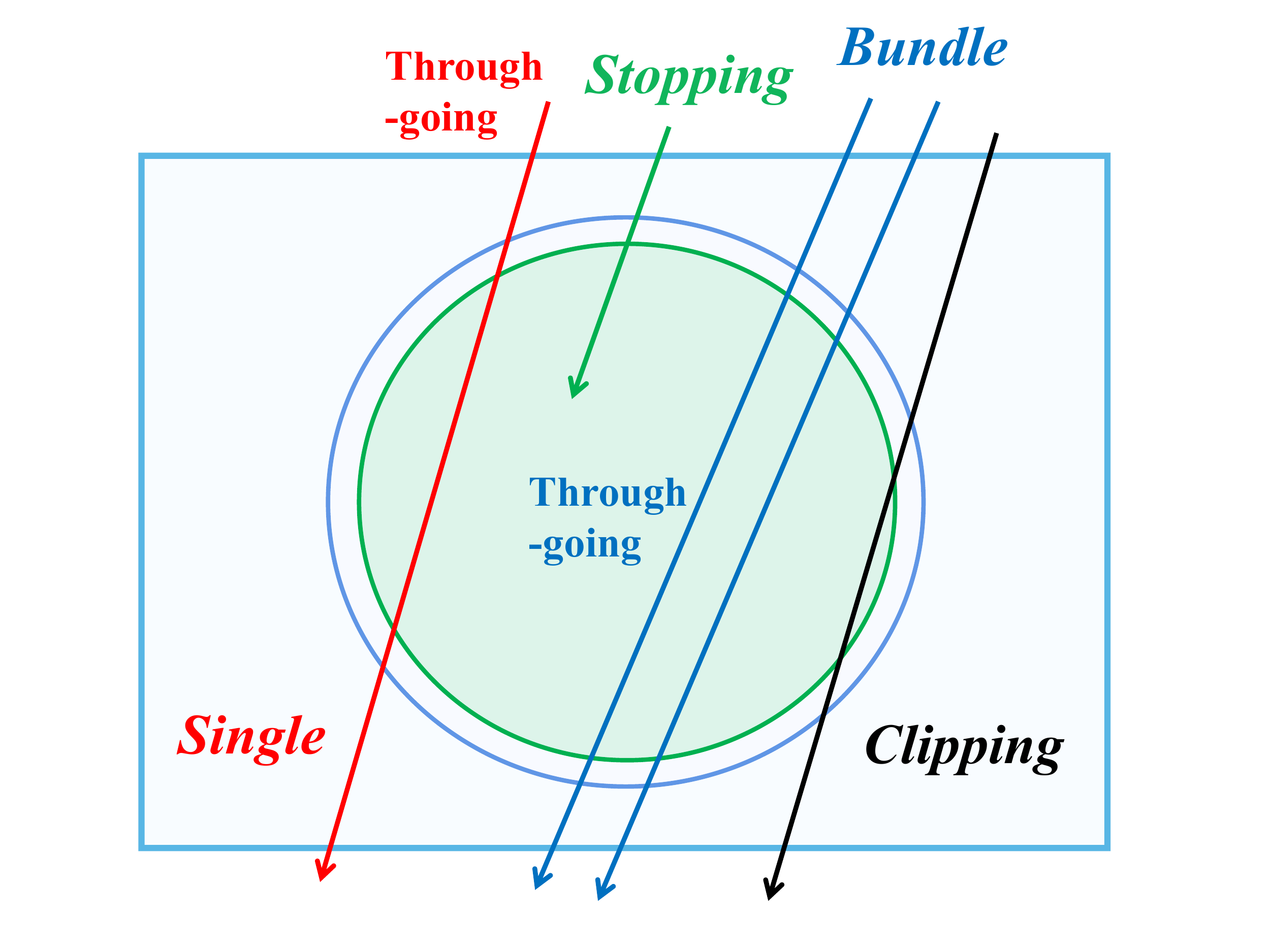}
		\caption{\label{fig: Muon_type} Schematic of a spherical target detector and different kinds of muon events labelled with single, bundle, through-going, stopping, and clipping.}
	\end{figure}

	In addition, based on muon multiplicity, muons can be classified into the following categories: single muon, double muons and multiple muons (Fig.~\ref{fig: Muon_type}). Muon bundles, corresponding to the muons whose multiplicity $\geq 2$, are remnants of air showers produced in the atmosphere by high energy cosmic ray nuclei~\cite{Gaisser}. When multiple muons go into an underground detector simultaneously, their tracks are approximately in parallel since they are produced in the same air shower, which typically occurs kilometers away from the detector \cite{MarcoGrassi}.  
	
	Three reconstruction algorithms have been developed for single muon, such as a method with a geometrical model which utilizes the geometrical shape of the fastest light~\cite{Muon_reconstruction_with_a_geometrical_model_in_JUNO}, a method with the fastest light model which utilizes the minimization of the first hit time (FHT)~\cite{the_fastest_light_in_the_JUNO_central_detector}, and new technology using deep learning and GPU acceleration~\cite{a_convolutional_neural_network}. But the reconstruction of muon bundles is not included and rarely discussed. The time needed for the first two algorithms is at the level of a few seconds. For those experiments with detectors of tens of meters, like JUNO and LENA~\cite{LENA}, etc, the proportion of muon bundles is larger, hence it is necessary to develop an algorithm to reconstruct multiple muon tracks. This paper proposes an efficient algorithm for the track reconstruction of both single and double muons that also demonstrates good reconstruction abilities for clipping muon based on the JUNO experiment. On the other hand, a few muon tracks are accompanied by electromagnetic or hadronic showers, called shower muon, and it is difficult to estimate where the shower takes place. A dedicated algorithm for shower muon is being developed, but it is not the subject of this article.
	
\section{The charge pattern of PMT hit array} 
\label{section:charge response of PMT}

\begin{figure*}[!htb]
        \centering
        \subfigure[Charge pattern of one single muon event.]{
            \includegraphics[width=0.45\textwidth]{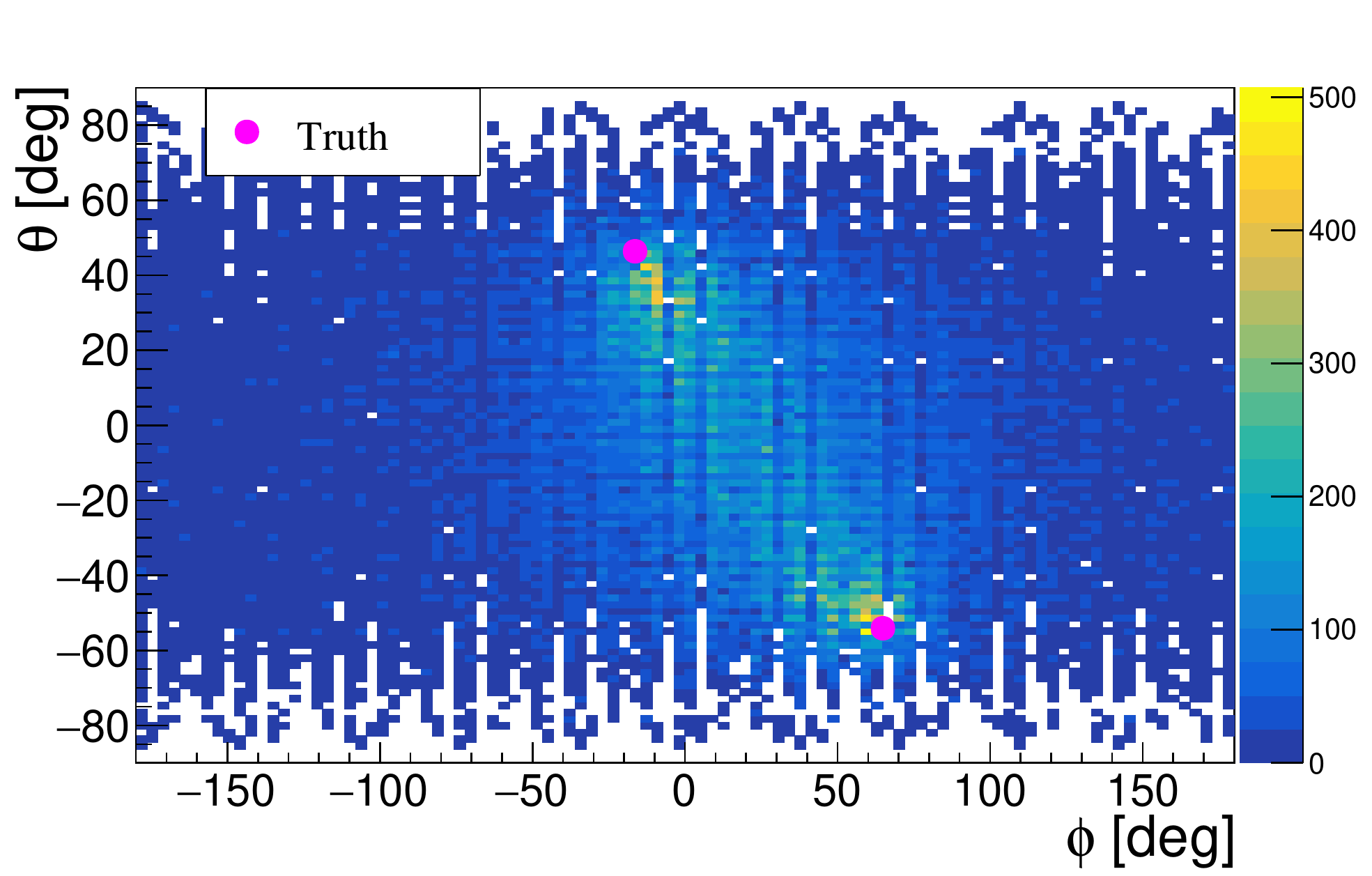}
            \label{hitpatterns_single}
        }
        \quad
        \subfigure[Charge pattern of one muon bundle event.]{
            \includegraphics[width=0.45\textwidth]{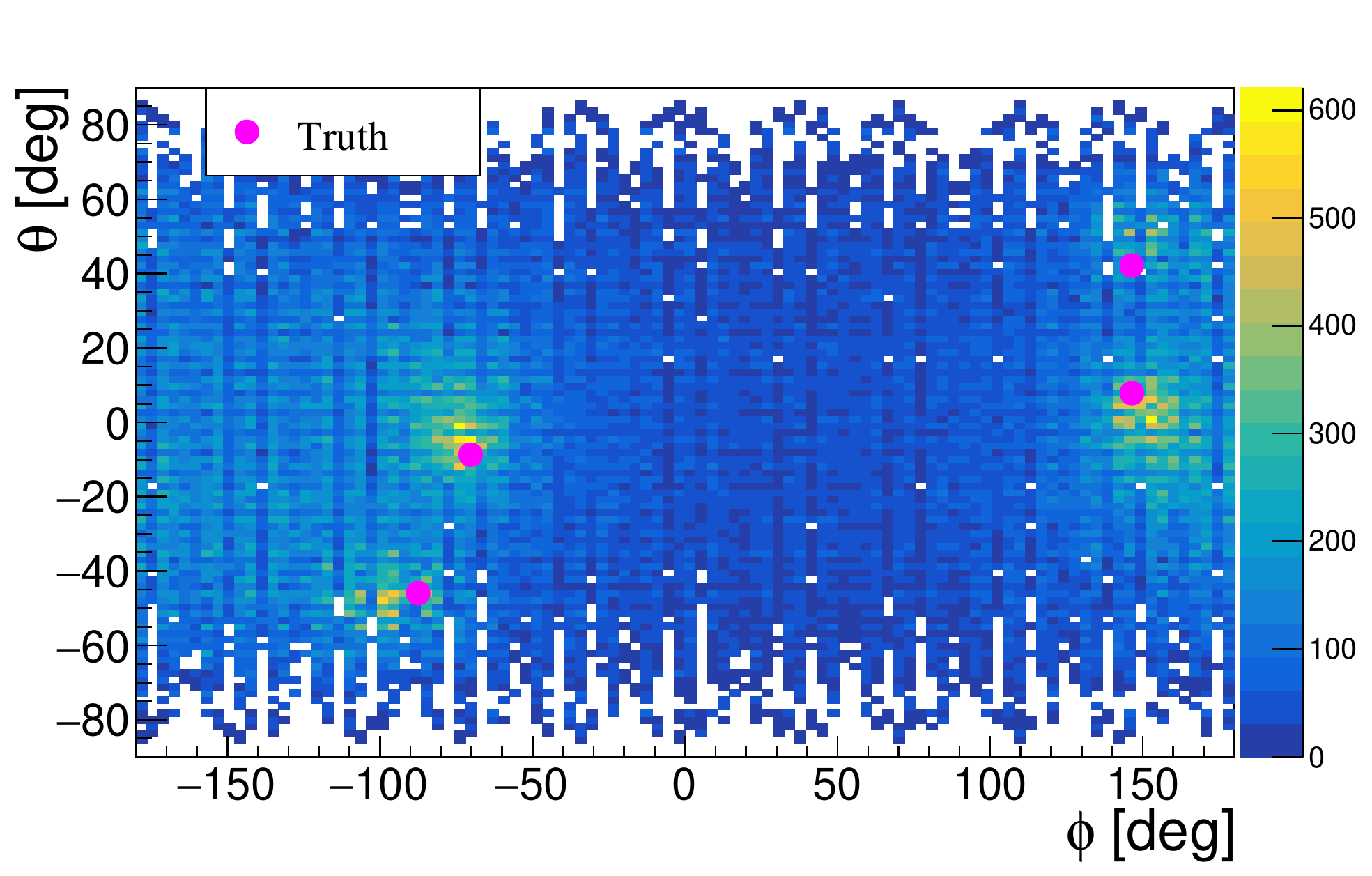}
            \label{hitpatterns_double}
        }
        \quad
        \subfigure[Charge pattern of one clipping muon event.]{
            \includegraphics[width=0.45\textwidth]{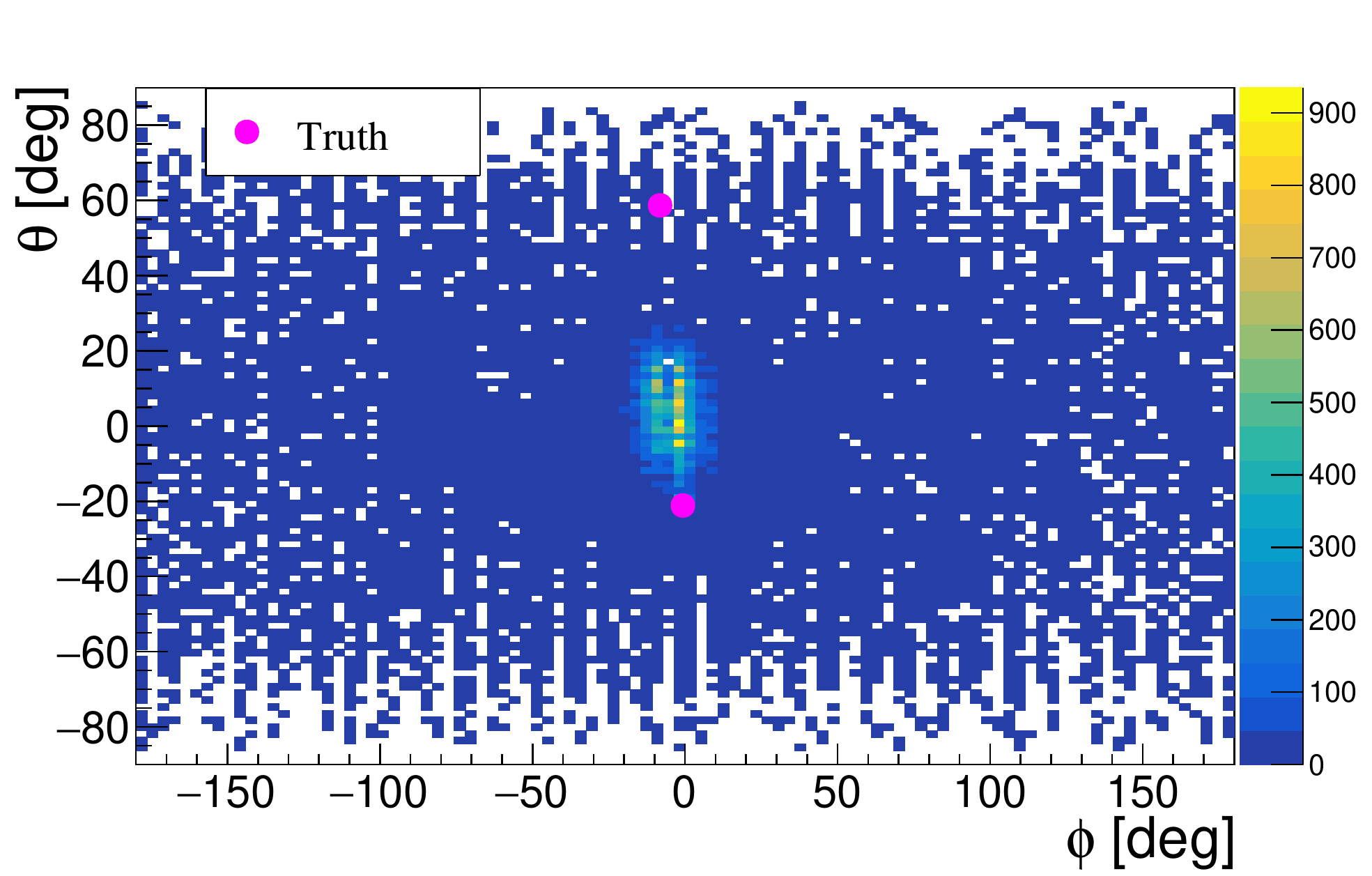}
            \label{hitpatterns_clipping}
        }
        \quad
        \subfigure[Charge pattern of one stopping muon event.]{
            \includegraphics[width=0.45\textwidth]{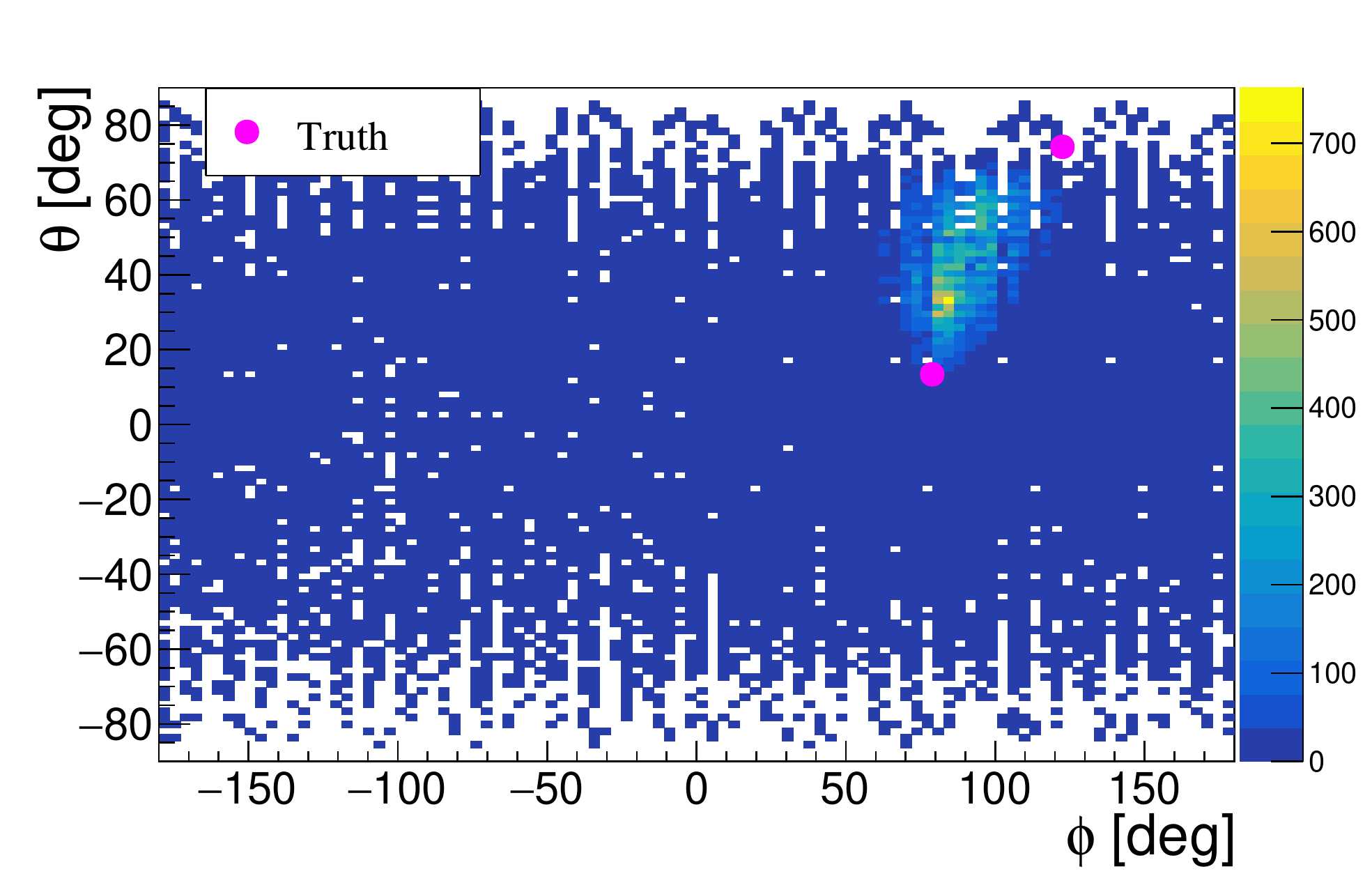}
            \label{hitpatterns_stop}
        }
        \caption{The charge pattern of the PMT hit array was projected on the $\theta-\phi$ 2D plane for the single, bundle, clipping, and stopping muons, respectively. The orange color indicates there are more than 400 PEs in the pixel. The purple solid dots correspond to the true incident and exit points of the muons. Since there is a water buffer with a thickness of 1.8 m between the PMT spherical shell and LS sphere, the charge cluster center in the charge pattern is not exactly equal to the muon true incident and exit points, more explanations can be found in Fig.~\ref{fig: Re-reconstruction}.}
        \label{hitpatterns_examples}
    \end{figure*}

	When a cosmic muon goes through the JUNO detector, the muon will leave track information in the detector, including time information and charge information. For example, we can see a PMT charge cluster at the incident point, and there is also another charge cluster at the exit point. To more easily deal with the charge clusters, the PMT hit array was projected on a two-dimension (2D) plane to show the charge pattern (Fig.~\ref{hitpatterns_examples}). In the charge pattern, the number of PMTs in each pixel varies from 0 to 8, depending on their locations.
	
	The fast light information around the muon track is widely used for single muon reconstruction, but it is more complicated to handle in the case of multiple muon tracks.
	To reconstruct muon bundle, the charge pattern of the PMT hit array was investigated and different kinds of muons can provide clear different charge clusters (Fig.~\ref{hitpatterns_examples}). The 20-inch PMTs near the muon incident point and the exit point will receive several thousands of photoelectrons (PEs), they are expected to be saturated and are not applied in this analysis. The 3-inch PMT (hereinafter referred to as PMT), with photocathode area about 40 times smaller than LPMT, work at a larger dynamic range and are proper for this study. 

	Different kinds of muons have different cluster features of the charge pattern. The cluster number has a relationship with muon track number and track behavior. The single through-going muon event has 2 clusters, but single clipping muon event only has 1 cluster since its incidence point and exit point are very close. The single stopping muon event only has 1 cluster since it only has one incident point but not an exit point. Double through-going muon event has 4 clusters, which are corresponding to two incident points and two exit points. There might be one or both tracks clipping in double muons, so the number of clusters is three or two. Similarly, the stopping double muon event includes 3 clusters (one muon stops) or 2 clusters (both muon stop). For muons whose multiplicity $\geq$ 3 (multiple muons), as effort to reconstruct multiple muons was yielding a low reconstruction efficiency and will not be further discussed in the paper.
	
\section {Reconstruction algorithm}
\label{section:Reconstruction algorithm}

\begin{figure*}[!htb]
		\centering
		\includegraphics[width=0.8\textwidth]{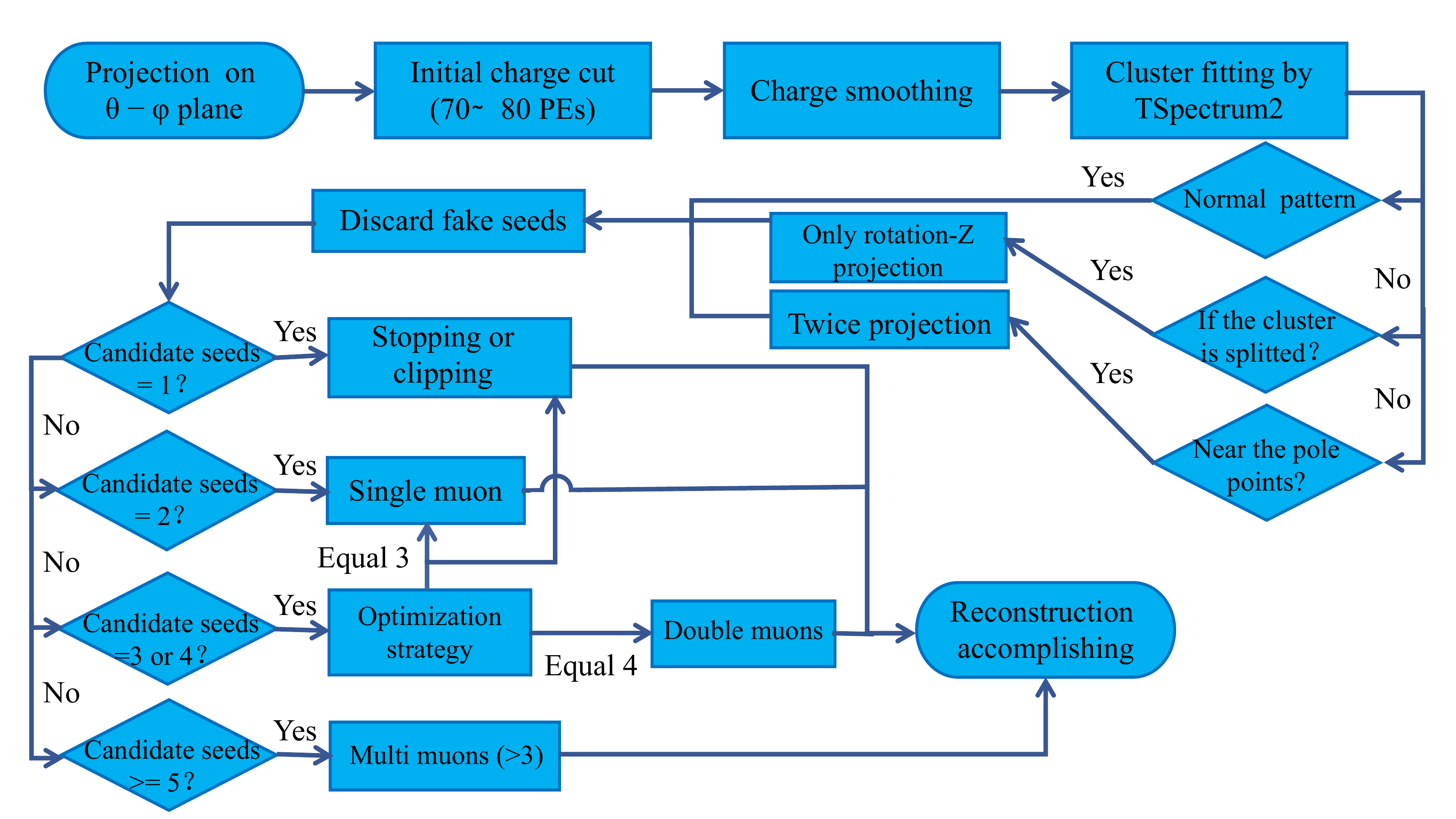}
		\caption{The flow chart of muon track reconstruction.} 
		\label{fig:flow-chart}
	\end{figure*}

	Using the information from the PMT charge pattern, a reconstruction algorithm was developed for both single muon and double muons (Fig.~\ref{fig:flow-chart}). The following is a brief step list of the algorithm flow, and more details will be introduced later.
	
	(1) Project the PMT charge pattern on the $\theta-\phi$ plane.
	
	(2) Discard PMTs whose charge is smaller than 75 PEs, more details in Fig.~\ref{fig:nPE_byPMT}. 
 
	(3) Apply charge smoothing method (section~\ref{Distortion elimination optimization}) to reduce the fake cluster finding when applying the fitting in the next step.
	
	(4) Fit with the ROOT tool `TSpectrum2'~\cite{root-tool} to find out the center positions of clusters (cluster seeds) in all the possible cluster candidates.
	
	(5) Rotate $90^\circ$ of PMT ball to let near North and South Pole's PMTs move to equator position (section~\ref{Distortion elimination optimization}), then repeat step (1) - (4), because there are many distortions when project PMTs from spherical surface to two-dimension plane, especially the area near North and South Pole.
	
	(6) Charge weighted calculation to correct each cluster seed and discard the fake seeds. 
 
	(7) Merge the adjacent candidate seeds if they are too close. 
	
    (8) Match two candidate seeds as a muon track. The additional strategy will be used to handle the case if there are more than two candidate seeds. More details can be found in section~\ref{Double muon optimization}. 
	
\subsection{Methodologies}
\label{subsection: Charge weight method}

	For the JUNO detector, the PMTs closely packed around the LS sphere make a PMT shell which is a 2 dimensions (2D) surface instead of 3 dimensions (3D), so the PMT charge hit pattern can be projected onto a 2D plane to do conveniently fitting. Like the world map of the Earth, there are several projection methods to get the 2D map, but every method introduces distortion through the projection. For the reconstruction, different projection methods get similar reconstruction results. Finally we selected the longitude-latitude projection method. The $\phi$ angle of PMT position is for X-axis on 2D plane and $\theta$ angle is for Y-axis.
	
\begin{figure}[!htb]
        \centering
        \includegraphics[width=0.45\textwidth]{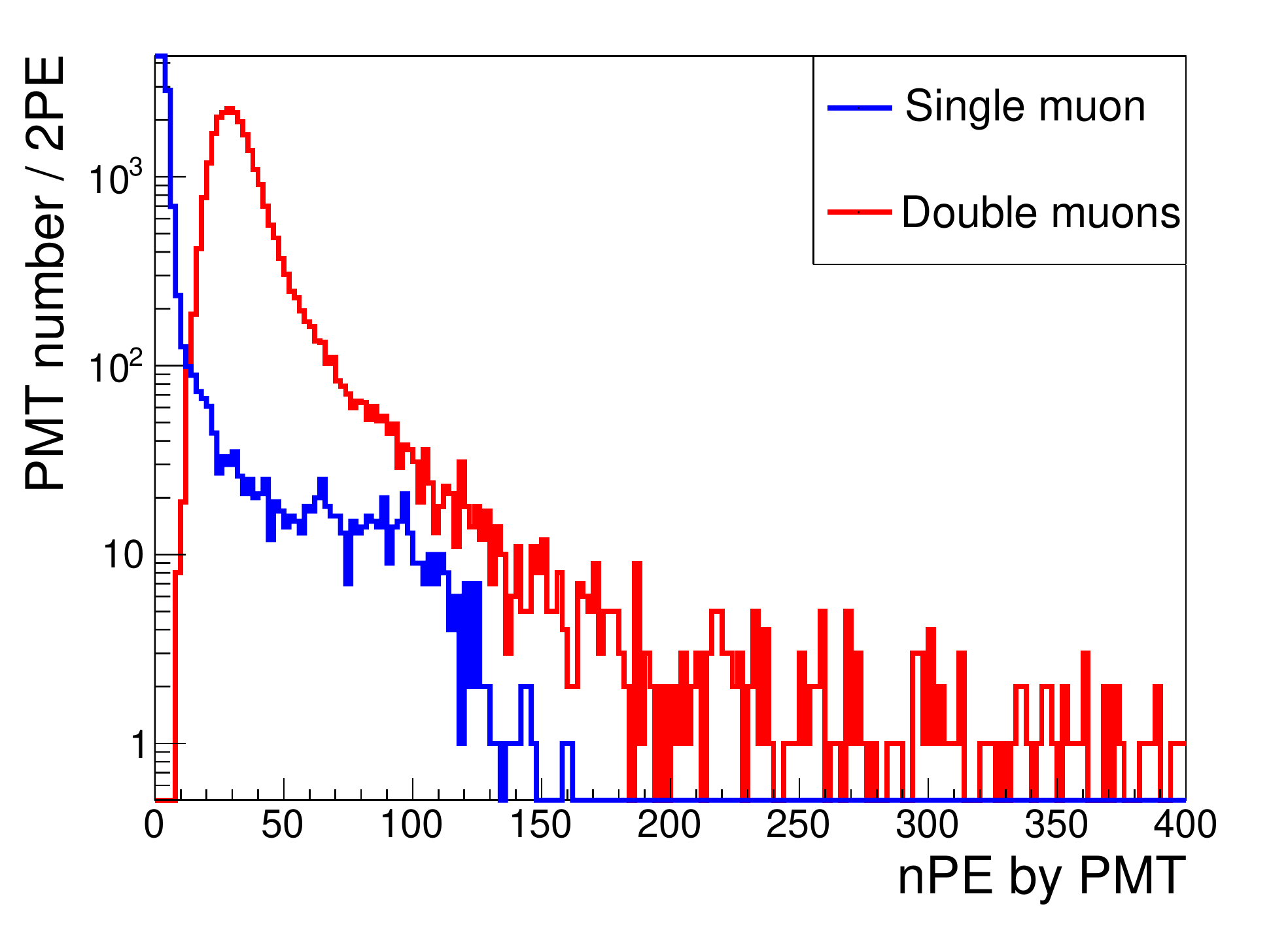}
		\caption{An example of the nPE distribution of a single through-going muon event (red) and a double through-going muon event (blue), respectively. The PMTs with a charge of more than 70 PEs generally correspond to the clusters caused by muon tracks. In order to highlight the clusters caused by muon incident and exit points, the PMTs whose charge is smaller than 75 PEs are discarded. } 
		\label{fig:nPE_byPMT}
    \end{figure}

	When a muon passes through LS, more than 90\% of photons are emitted by the scintillation process which emits light isotropically. The closest PMTs from muon incident point or exit point will collect the most PEs, which will make two clusters on the PMT charge hit pattern (Fig.~\ref{hitpatterns_single}). Fig.~\ref{fig:nPE_byPMT} is an example of the nPE distribution of a single through-going muon event and a double through-going muon event, respectively. The PMTs with a charge of more than 70 PEs generally correspond to the clusters caused by muon tracks. To highlight the clusters which were caused by muon incident and exit points, we eliminate the PMT noise ($\sim$ 1~PE), radioactivity (several PEs) and PMTs far from clusters (tens of PEs), the PMTs whose charges are smaller than 75~PEs are discarded. In the simulation study, when the maximum discard charge varied from 70~PEs to 80~PEs, the reconstruction result was not significantly changed, more details could be found later (section~\ref{subsection:veto efficiency of reconstructed muon track} and Fig.~\ref{fig:PECUT}).

\begin{figure*}[!htb]
        \centering
        \subfigure[]{
            \includegraphics[width=0.45\textwidth]{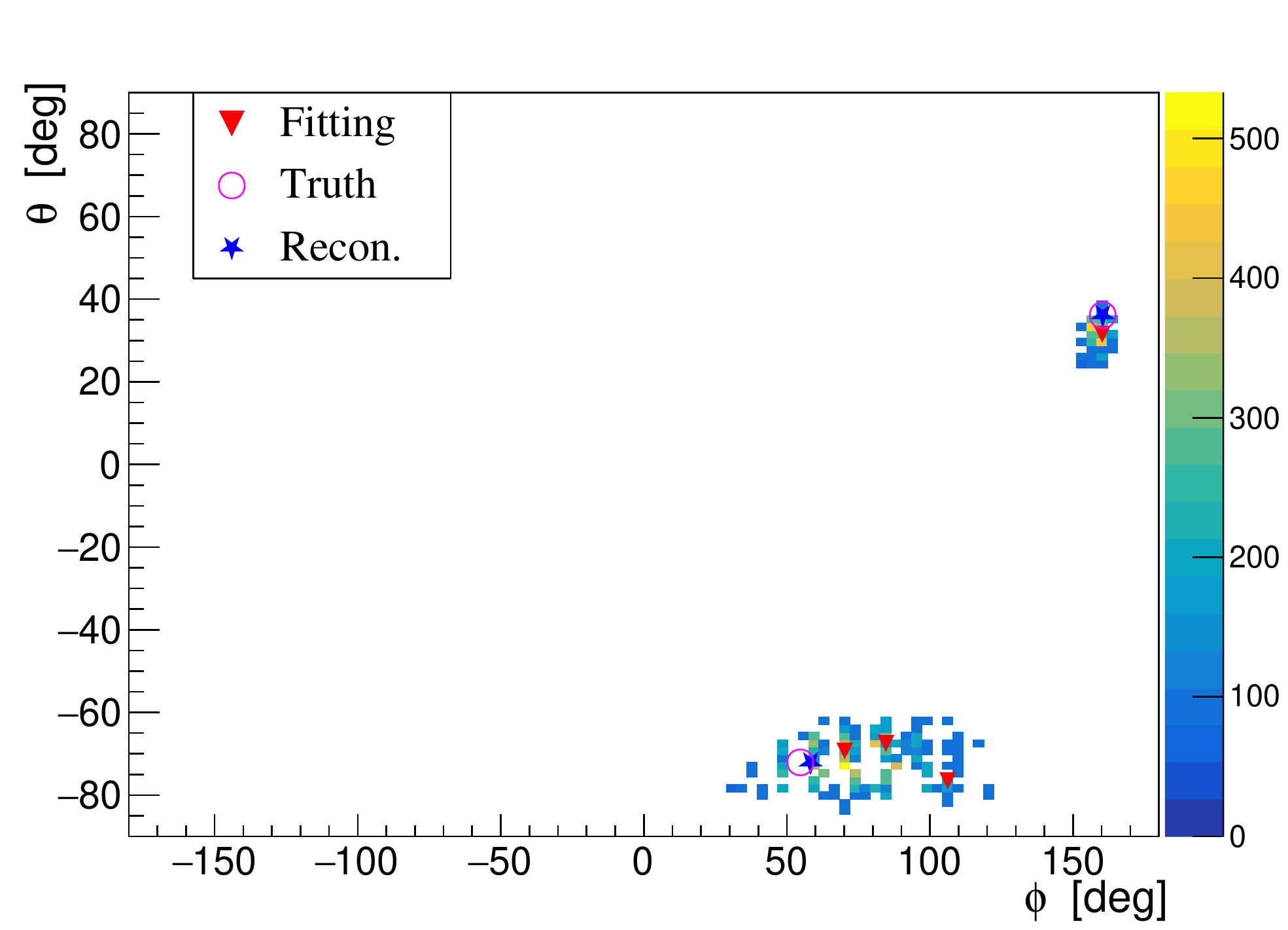}
            \label{fig: Before Charge smoothing}
        }
        \quad
        \subfigure[]{
            \includegraphics[width=0.45\textwidth]{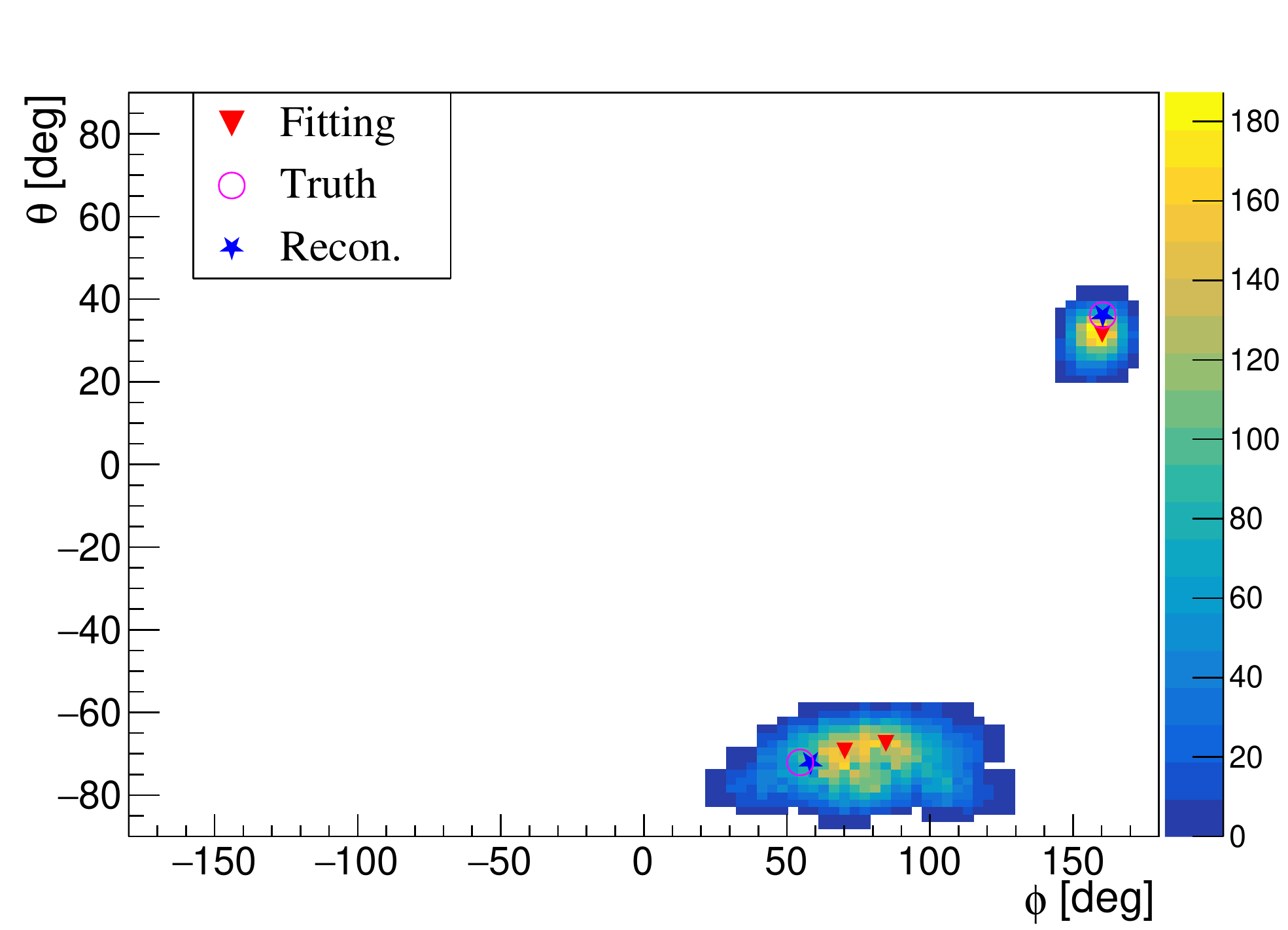}
            \label{fig: After Charge smoothing}
        }
        \caption{The hit patterns of a single through-going muon event before and after smoothing. (a) The PMTs whose charge are smaller than 75 PEs are discarded, as a result, the pixels only with little charges are not shown. The red triangles are the cluster seeds fitted by `TSpectrum2'. Two purple circles are the muon true incident and exit points on the pattern. The blue stars are the final reconstructed muon track incident and exit points. (b)  The smoothed charge cluster. The same labels as in Fig.~\ref{fig: Before Charge smoothing} are used. The cluster seeds (red triangles) were reduced from three to two for the bottom cluster. }
        \label{hitpatterns}
    \end{figure*}

	On the PMT charge pattern, a 2D peak searching and fitting algorithm named `TSpectrum2' was investigated and used. `TSpectrum2' is a ROOT tool~\cite{root-tool}, which is based on a two-dimensional Gaussian function fitting and it searches for peaks in source spectrum by deconvolution method. Its goal was to get the maximum local peak by calculating the spectrum of the standard deviations of the smoothed second derivatives in two dimensions~\cite{Identification_of_peaks}. On success it returns the number of found peaks and their locations. The order of found peaks is arranged according to their heights in the spectrum. In our study, the maximum 7 clusters was set to only list the highest 7 peaks which were enough for single muon (2 peaks), double muons (4 peaks) and triple muons (6 peaks) searching. The tracking precision of triple muons was much lower than double muons, so it was not shown in this paper, but we still set 7 and kept the reconstruction potential for triple muons by this method and it will be studied further in the future. The function `Search()' was used to do the fitting~\cite{root-tool} and the peak parameter `sigma' was tuned as 2, which can be set from 1 to 10. During the tuning, it showed the results were changed not too much, which means the function is robust. The parameter `threshold' was set to 0.05, which means do not fit the small peaks whose amplitude was less than 5\% of maximum peaks. The tool runs very fast and can mathematically find out the cluster seed candidates (Fig.~\ref{fig: After Charge smoothing}).   
	
	However, the fitting may find more seeds that can not match the muon incident and exit points and caused by the charge statistical fluctuation of the PMT charge. The collected charges of adjacent PMTs surrounding the largest charge collection PMT is not a continuous drop from the cluster center to the cluster edge. On the other hand, some muons may have a relatively large energy deposit at a certain position along the track. All of these may result in the wrong cluster number and fake cluster seeds when matching with true muon hit points. Fig.~\ref{fig: Before Charge smoothing} shows an example of the PMT charge hit pattern of a single through-going muon event before charge smoothing above the PMT charge threshold. It shows that the fitting found three cluster seeds near the muon exit point where we only need one cluster seed. The fitting results of cluster seeds need to be carefully checked and analyzed.
	
	To discard the fake cluster seeds and get the muon true hit points, a charge weighted algorithm was developed. Firstly, based on each cluster seed, we calculate the charge weighted position of fired PMTs in a four meter radius with the formula:
	\begin{large}
		\begin{equation} 
		\bar x =\frac{\sum\limits_{i=0}^{n}q_{i}x_{i}}{\sum\limits_{i=0}^{n}q_{i}}
		\end{equation} 
	\end{large}
	 
	Where $x_{i}$ is the spatial position of PMT$_{i}$; $n$ is the fired PMT number in a four meter radius around the cluster seed; $q_{i}$ is the charge collected by PMT$_{i}$. If the total charge in a four meter radius around the cluster seed is less than 1000~PEs, the cluster seed will be discarded

	Secondly, after calculation, replace the cluster seed by the charge weighted position ($\bar x$). Then, if two new cluster seeds are too close ($<$ 4~m), they will be merged into one point by their charge weights. This step can filter out most of the fake seeds given by `TSpectrum2' fitting and give a correct position. Thirdly, if the gathered cluster charge within a radius of four meters around the cluster seed is too small ($<$ 10\% of total charge of all clusters), the seed also will be discarded. This can eliminate the small charge clusters found by `TSpectrum2' fitting which can not match with muon true incident and exit points. Finally, the charge weighted position of the remaining clusters are considered as the muon incident or exit point.
	
	For the single muon event, the two cluster seeds can be easily matched with the muon track and give the cylindrical veto cut along the muon track. Even if there is only one cluster seed, we also can take the seed as center point to provide the spherical veto with radius 7~m (tuned value). But it is complex for a muon bundle event, there we will have 4 or 3 reconstructed points to match with the two tracks. A dedicated algorithm was developed and introduced in section~\ref{Double muon optimization}. 
	
\subsection{Distortion elimination and performance optimization}
\label{Distortion elimination optimization}

	To make the muon track reconstruction more precise, we eliminate distortion by the rotation method, remove the statistical fluctuation of charge cluster by smoothing method and transfer the cluster seeds on PMT charge pattern onto the true muon track.
	
	As previously mentioned that projection method (normal projection) brings non-negligible distortion especially near the polar zone. To avoid the projection distortion, we rotate the PMT shell $90^\circ$ around X-axis to let North and South pole move to equator (Fig.~\ref{fig:  rotation method of PMTs}), then do the projection again (rotation-X projection). The cluster seeds in the $\theta$ range ($-50^\circ$, $50^\circ$) will be selected in the normal projection and cluster seeds in the original $\theta$ range $>$~$40^\circ$ or $<$~$-40^\circ$ will be selected in the rotation-X projection. Some repeated seeds in the overlap area ($40^\circ$, $50^\circ$) and ($-50^\circ$, $-40^\circ$) will be merged during the charge weighted step. 
	
	\begin{figure}[ht]
		\centering
		\includegraphics[width=0.6\textwidth]{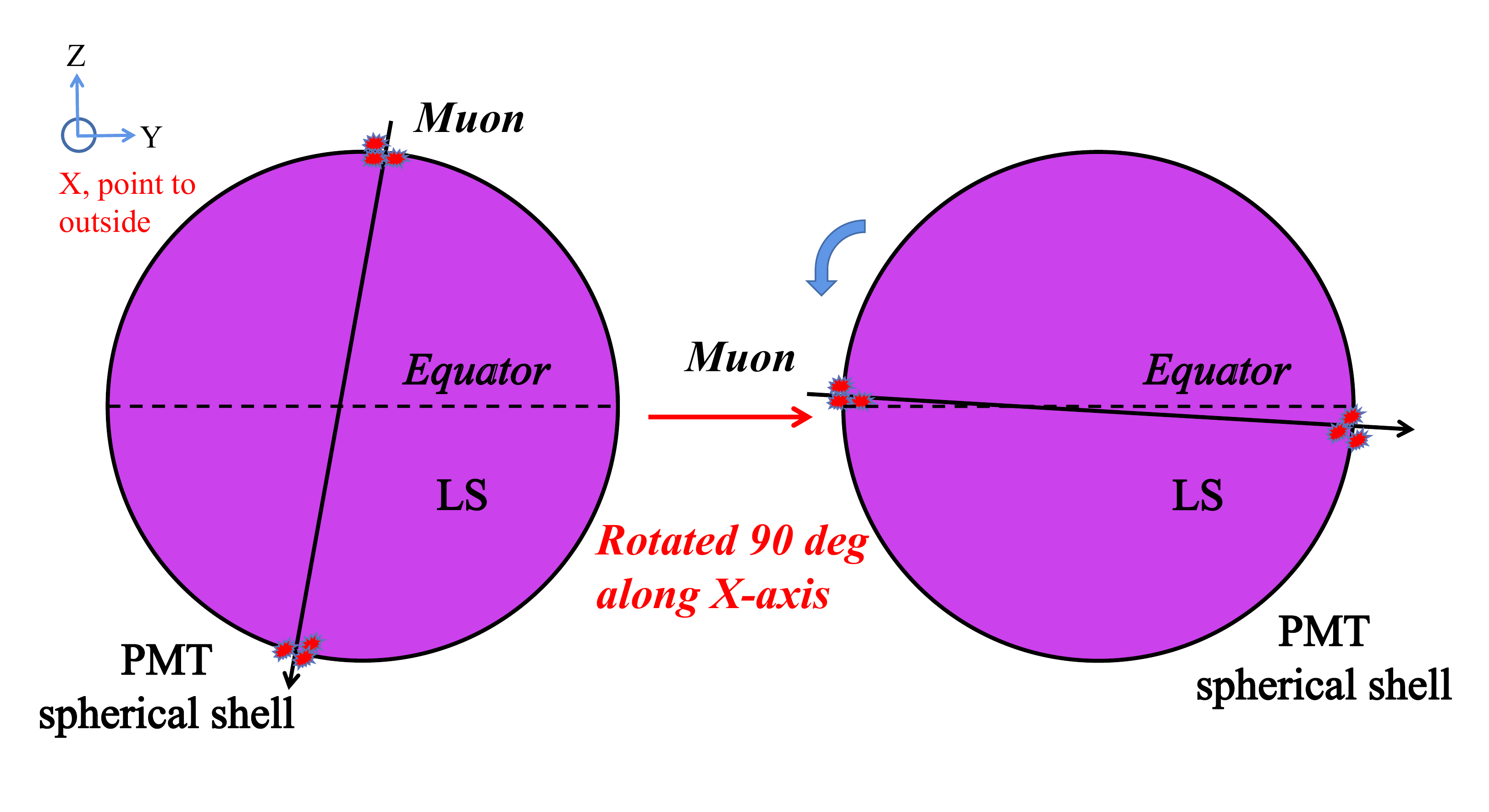}
		\caption{The rotation method of PMTs to reduce the distortion of projection.}\label{fig: rotation method of PMTs}
	\end{figure} 
	
	Sometimes some clusters are split from one cluster into two clusters when incident or exit points are located at map's edge ($\phi$ angle is over $160^\circ$ or less $-160^\circ$). At first, the PMTs will be rotated $90^\circ$ around the Z-axis (rotation-Z projection) to get the precise position of clusters that are located at the edge of 2D charge spectrum. And then combining with normal projection and Z-projection, we will check whether some fitted points are located at the edge or not. If a cluster was split into two parts, fitting result from rotation-Z projection will be adopted.
	
    In addition, there is another uncertainty. Due to the statistical fluctuation of the PE number collected by PMT or emitted by LS and maybe a relatively large energy deposit at a certain place along the muon track, there are not only the peaks induced by injection and exit points, but also some relatively small peaks. Fig.~\ref{fig: Before Charge smoothing} shows all reconstructed peaks. Meanwhile, the `TSpectrum2' tool maybe supplies fitted points at the non-hit position. To remove the uncertainty element, each pixel's charge was smoothed ( $q^{s}$) based on the adjacent pixel charges with the smoothing formula:
	\begin{large} 
		\begin{equation} 
		%q^{smooth}_{n} = \frac{\sum^{m}_{1}(\frac{deg_{0}}{deg_{1}+deg_{m}}q_{ni})}{\sum^{m}_{1}(\frac{deg_{0}}{deg_{1}+deg_{m}})}\\
		q^{s} = \frac{\sum\limits_{i=0}^{N}(\frac{w_{0}}{w_{1}+d_{i}}q_{i})}{\sum\limits_{i=0}^{N}(\frac{w_{0}}{w_{1}+d_{i}})}
		\end{equation} 
	\end{large}
	
	Where $q_{i}$ is the charge of the $i$-th adjacent pixel; $w_{0}$ and $w_{1}$ are the weight factors and the best tuning result is $w_{0}/w_{1}$ = 4/9; $d_{i}$ is the distance factor between the smoothing pixel and the $i$-th adjacent pixel. The distance factor of the nearest pixel is 1. The further of pixel distance, the larger of $d_{i}$ and the lower weight of $q_{i}$. $N$ is the number of the adjacent pixels that are used for charge smoothing and surround the specified pixel, which is set to 24. The comparison of Fig.~\ref{fig: Before Charge smoothing} and Fig.~\ref{fig: After Charge smoothing} shows that the charge smoothing method can  make the cluster peaks more protruding and reduced the number of false fitting peaks which obviously improved the reconstruction efficiency. 
	
\begin{figure}[!htb]
		\centering
		\includegraphics[width=0.45\textwidth]{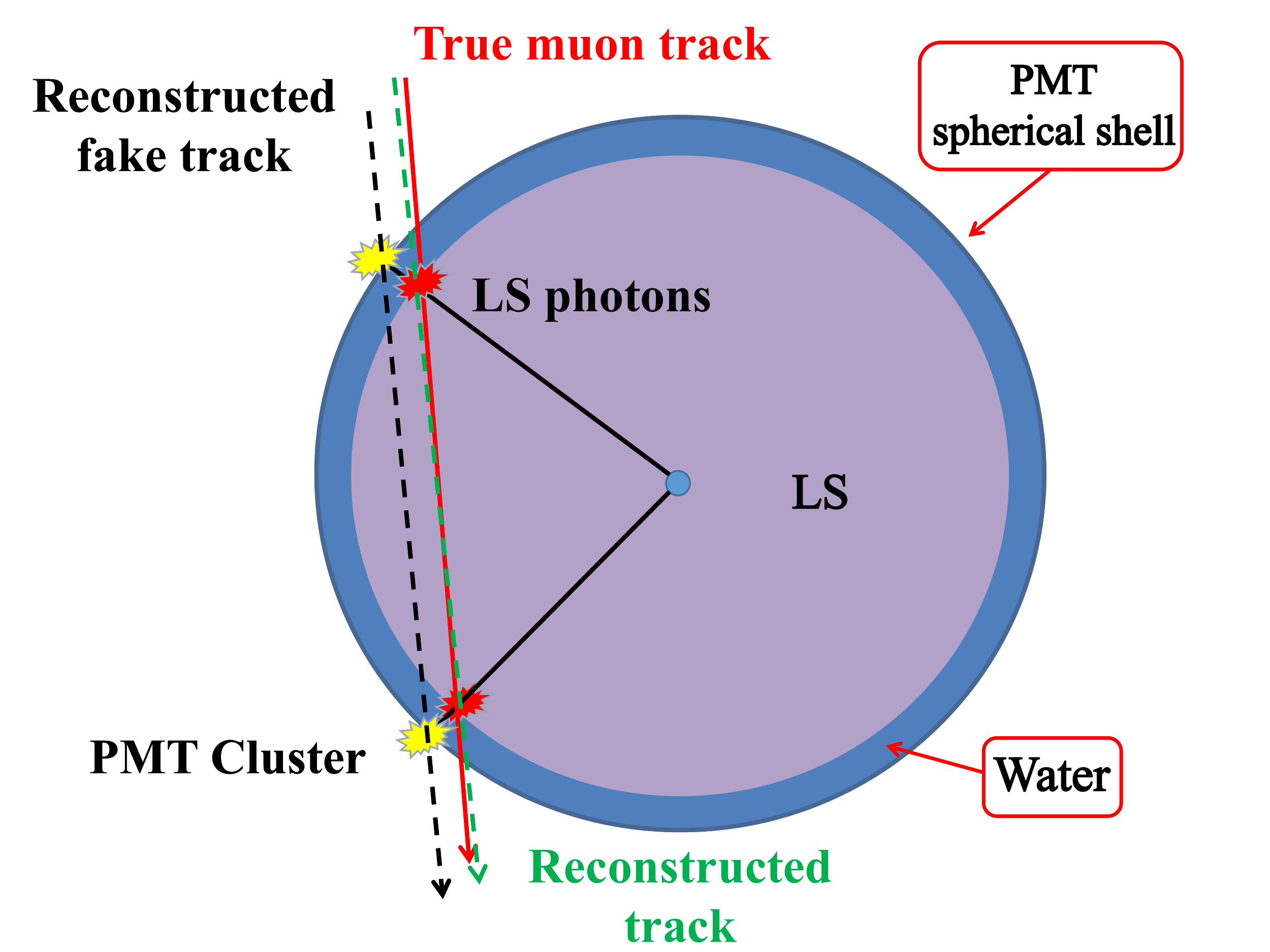}
		\caption{\label{fig: Re-reconstruction} The schematic picture of a single muon track reconstruction with two clusters. In JUNO detector, there is a water buffer with a thickness of 1.8~m between the PMT spherical shell and LS sphere. The cluster charge centers (yellow clusters) make a fake track and need be corrected by geometric effect to get the reconstructed track.}
	\end{figure}

    Actually, the charge cluster center is not exactly equal to the muon real incident and exit point. As Fig.~\ref{fig: Re-reconstruction} shows that, there is a water buffer with a thickness of 1.8~m between the PMT spherical shell and LS sphere. When a muon goes in the PMT shell, most photons come from LS, PMT charge cluster is close to the muon hit point of LS, so the reconstructed track just by the PMT charge cluster center is not the true muon track. The cluster charge centers (yellow clusters in Fig.~\ref{fig: Re-reconstruction}) have to be translated into the true muon hit points on LS sphere (red clusters in Fig.~\ref{fig: Re-reconstruction}) according to their spatial positions. Finally, we can reconstruct the muon track (green dotted line) by the reconstructed incident and exit points.
    
\subsection{Double muon event reconstruction strategies}
\label{Double muon optimization}

\begin{figure*}[!htb]
		\centering
		\includegraphics[width=0.9\textwidth]{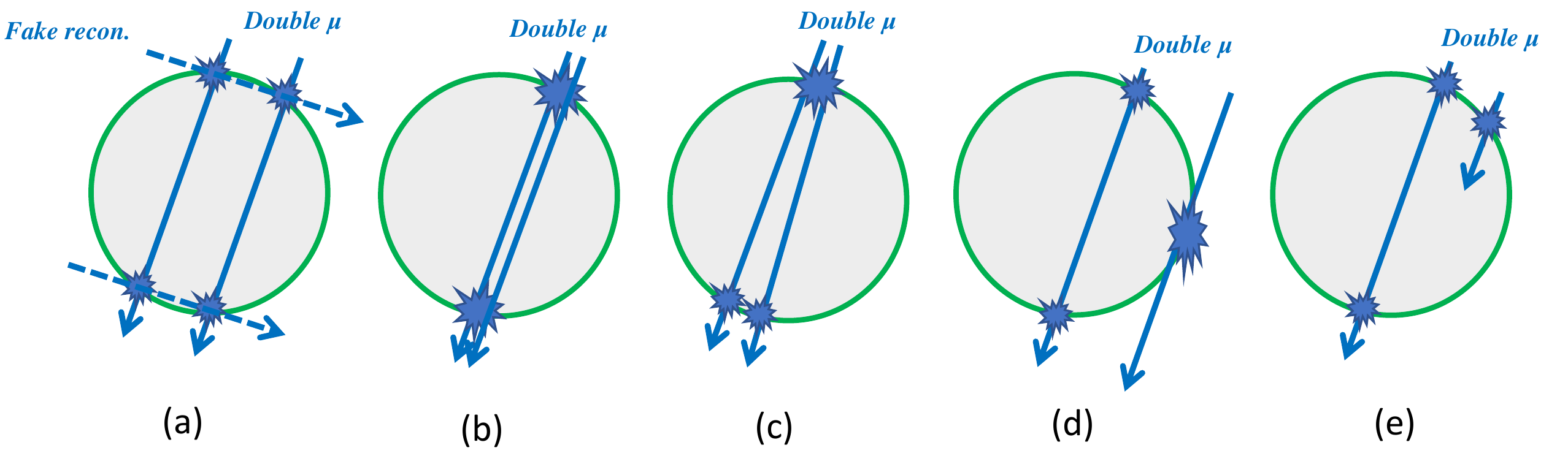}
		\caption{The schematic plots of different double muon events hit CD and their clusters. (a) Normal double muon event has two parallel tracks, hits detector almost at the same time and has clear 4 clusters. (b) Two close tracks hit CD have two clusters because the cluster overlap. (c) Two relatively close ($\sim$~4~m) tracks with a small angle ($<$~$2^\circ$) the two incident points have one cluster and the two exit points have 2 clusters. (d) One clipping track in double muons and 3 clusters. (e) One-stop muon in double muons and 3 clusters.} 
		\label{fig:double-muons}
	\end{figure*}

	As introduced in table~\ref{tab-class}, double muons usually have 4 hit points. Every two points compose one track candidate. There are three combinations from four hit points, and two of them are in parallel thus can be considered as the muon bundle candidates, as shown in Fig.~\ref{fig:double-muons}(a). The PMT timing information is used to further identify the true muon tracks.  
	
	With each reconstructed point in the center, the first hit time of their surrounding PMTs within 4~m are averaged as reconstructed time. The two earlier times are treated as two incident points, and the other two points are treated as the two exit points. This time strategy can guide and assist the correct track matching.     
	
	Generally, the normal cluster size is about 4~m. If two tracks are too close to each other, only two cluster seeds can be found (Fig.~\ref{fig:double-muons}(b)). In this case, the two cluster seeds will be connected and reconstructed into one single track. Although two tracks are close, the overlapped cluster is larger than one normal cluster. The distance of two muon tracks can be deduced by the overlapped cluster information. The primary requirement of the muon track reconstruction is to veto muon induced isotopes through a cylindrical volume cut along the muon track. Although only one track is reconstructed, the cylindrical cut with larger radius will be taken into account and can give the same background reduction ability.
	
	If double muons only have three charge clusters, it is not easy to distinguish between two muon tracks. In the study, we found there are two cases for a three cluster situation. One case is when two incident points overlap with each other (Fig.~\ref{fig:double-muons}(c)). We will calculate the averaged first hit time ($\bar T$) of surrounding PMTs within 4~m from each cluster. Because two tracks in double muons shoot in LS at the same time, the $\bar T$ of inject or exit points are close. In three values of $\bar T$, if the two $\bar T$ difference is less than 5~ns, both of these two clusters would be recognized as exit points and the other one is the incident point. Then two muons can be correctly tracked by using the incident point twice. Although the incident point is not very precise, the larger radius cylindrical cut according incident cluster size along each track still can give the same background reduction of muon induced isotopes.
	
\begin{figure*}[!htb]
        \centering
        \subfigure[An example of the three clusters (A, B and C) charge pattern by a double muon event which have one edge muon track.]{
            \includegraphics[width=0.45\textwidth]{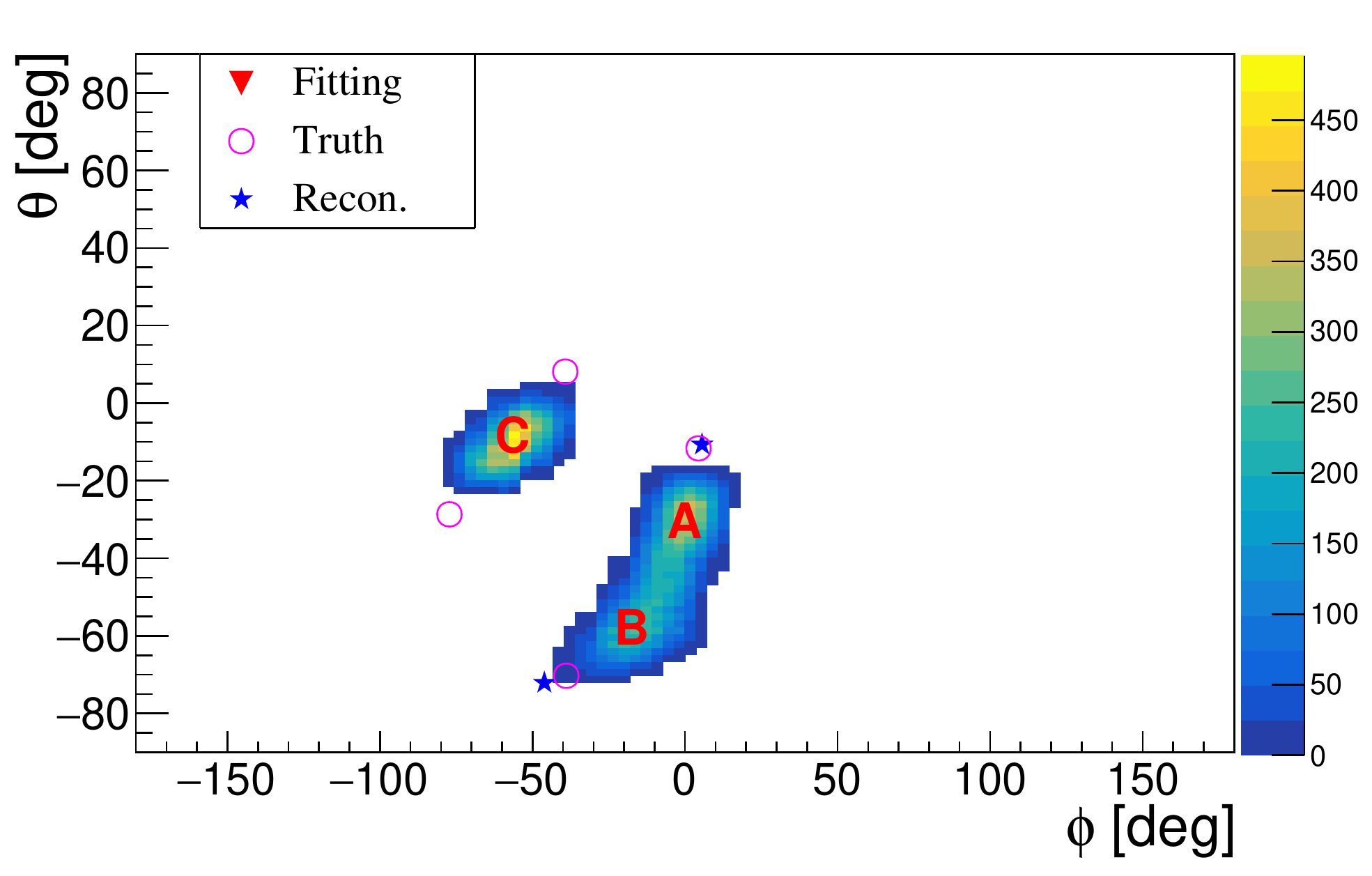}
            \label{fig: Three_Clusters_D}
        }
        \quad
        \subfigure[An example of the three cluster charge pattern by a single muon event whose distance to the CD center is about 14~m and near the edge of the CD.]{
            \includegraphics[width=0.45\textwidth]{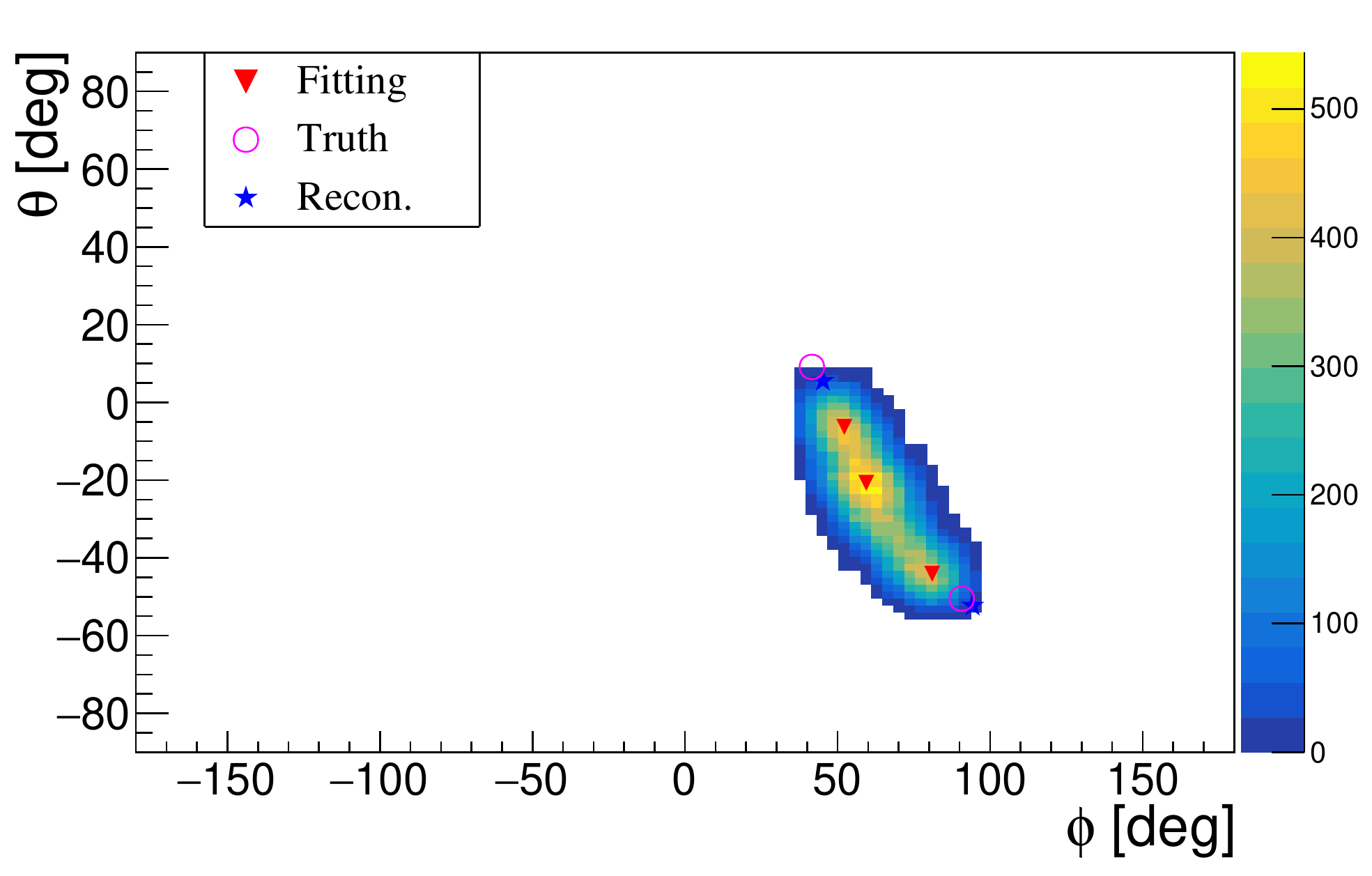}
            \label{fig: Three_Clusters_S}
        }
        \caption{An example of the three cluster charge patterns from double muon and single muon.}
    \end{figure*}

	The other case is that one clipping (Fig.~\ref{fig:double-muons}(d)) or stops in the CD (Fig.~\ref{fig:double-muons}(e)), which also produces three clusters. Projection on the 2D charge histogram, shown in Fig.~\ref{fig: Three_Clusters_D} as an example. Although each two clusters' time ($\bar T$) difference is larger than 5~ns, we still need to select two seeds to make a track. A direction judgment method was developed on the 2D cluster plot. When a muon track in LS is further from CD center, such as clipping muon, the PMT charge cluster is shaped like an ellipsoid. The direction of the ellipse major axis is along muon track direction. 
	The major axis directions of the ellipse can be expressed by vector $\overrightarrow{V_{A}}$, $\overrightarrow{V_{B}}$, $\overrightarrow{V_{C}}$ for the three clusters respectively (Fig.~\ref{fig: Three_Clusters_D}). We can calculate the angles of the candidate track ($L_{AB}$, $L_{BC}$ or $L_{AC}$) with its corresponding cluster vectors as the equations:
	
    \begin{equation}
        \begin{split}
        \angle~1 = \angle~\overrightarrow{V_{A}} \& L_{AB} +  \angle~\overrightarrow{V_{B}} \& L_{AB} \\
        \angle~2 = \angle~\overrightarrow{V_{B}} \& L_{BC} +  \angle~\overrightarrow{V_{C}} \& L_{BC} \\
        \angle~3 = \angle~\overrightarrow{V_{A}} \& L_{AC} +  \angle~\overrightarrow{V_{C}} \& L_{AC}
        \end{split}
        \label{eq:3angles}
    \end{equation}
    
	Where $\angle~\overrightarrow{V_{A}} \& L_{AB}$ is the acute angle of $\overrightarrow{V_{A}}$ and line $L_{AB}$, so are the other angles. The minimum one in three angles $\alpha_{1}$, $\alpha_{2}$ and $\alpha_{3}$ will be selected, and the corresponding two clusters will be set to one muon track and the line direction will be selected as the muon direction.
	
    The single muon event may also have three clusters which may confuse the double muon reconstruction. When a single muon track is close to the edge of LS sphere (the spatial distance between muon track and LS center is from 13~m to 16~m), the photons emission on muon track all are not far from PMTs and PMTs almost all fired (Fig.~\ref{fig: Three_Clusters_S}). 
    For this case, we still calculate the corresponding three angle values by equation~\ref{eq:3angles}. The feature is all three angles are less than $5^\circ$. So we use this feature to distinguish the single muon or the double muons. The middle point (Fig.~\ref{fig: Three_Clusters_S}) is a false point and the remaining two points are selected as the inject and exit points.
	
\section {Reconstruction performance and discussion} 
\label{section:Reconstruction performance and discussion}

    To optimize and validate the reconstruction algorithm in this paper, about 380 thousand simulated muon samples were produced with the JUNO offline software~\cite{Li_2017zku,Li:2018fny}. In the simulation, the cosmic air shower simulation in the JUNO site, the local geological map and the details of detector structure have been considered, resulting in the muon rate at about 0.004~Hz/m${^2}$ (3.4$\times$ 10$^5$ events/day) with average muon energy of about 207~GeV~\cite{JUNO-PPNP}. The muon samples include $\sim$92\% single muon events, $\sim$6\% double muon events, and less than 2\% of other events whose muon multiplicity $\geq$ 3. The muon energy deposit range is from 0 to $\sim$100~GeV. 
	For low background experiments, like neutrino experiments and dark matter experiments, muon-induced backgrounds suppression is the main motivation of muon reconstruction. The reconstruction resolution and accuracy of the muon track is the basis for muon veto strategy and backgrounds rejection. In addition, the reconstruction efficiency, robustness and speed should also be considered to evaluate the performances of the reconstruction algorithm, especially for high muon rate and large volume detector experiments. 
	
\subsection{Reconstruction performances of single muon and double muons}
\label{subsection:Reconstruction performances of single and double muons}

\begin{figure}[!htb]
		\centering
		\includegraphics[width=.45\textwidth]{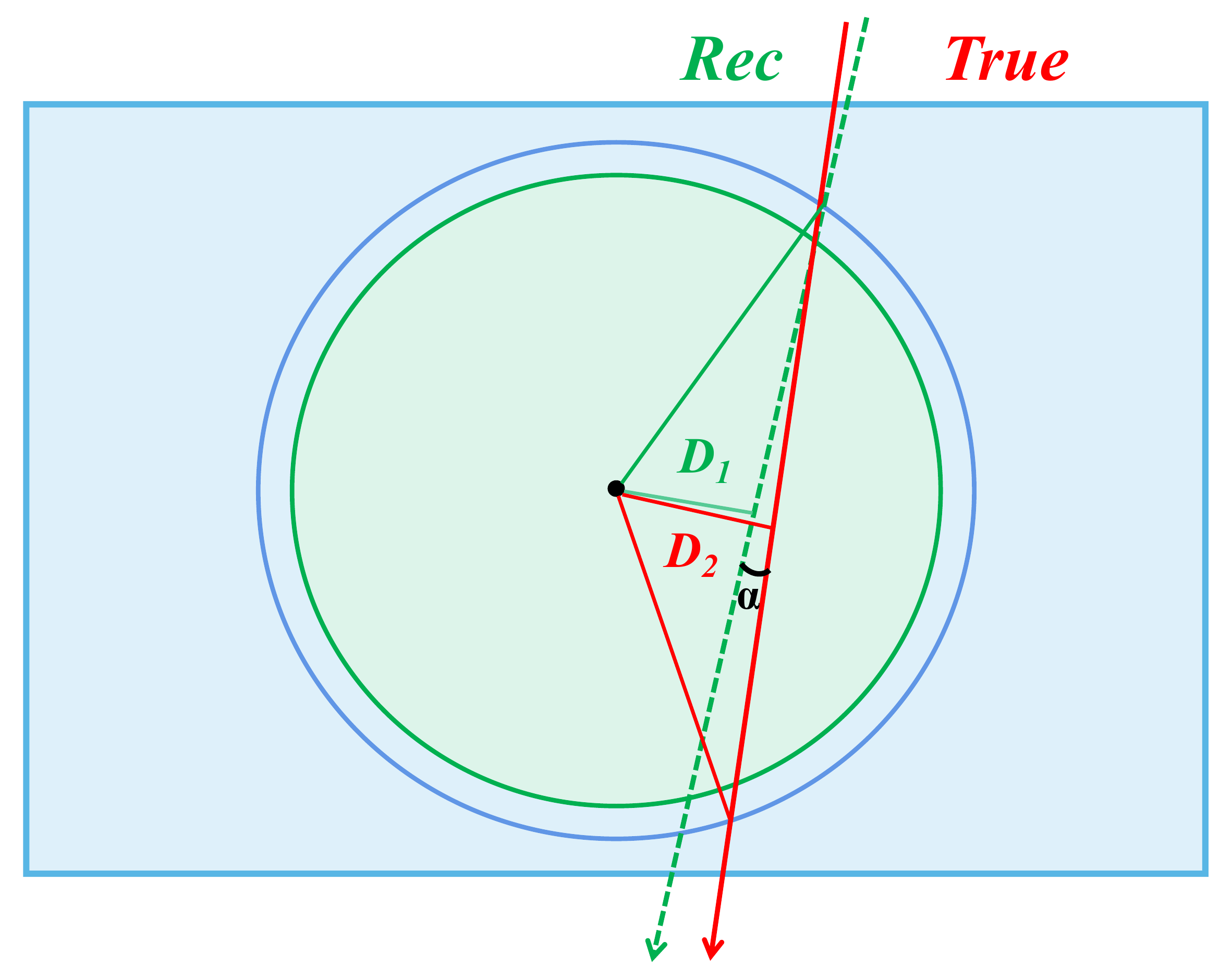}
		\caption{The schematic of parameters $\Delta D$ and $\alpha$. $\Delta D = D_{1}- D_{2}$. $D_{1}$ and $D_{2}$ are the distance from the CD center to the reconstructed track and the true track respectively, and $\alpha$ is the acute angle between the reconstructed track and the true track.} 
		\label{fig:rec-parameters}
	\end{figure}

    To obtain more exposure for the signal (neutrinos or others), the basic veto strategy is to apply the partly volume veto along the muon track for a period of time. The relative position of the reconstructed track and true track can be described using the following two parameters: $\Delta D$ and $\alpha$ (Fig.~\ref{fig:rec-parameters}). $\Delta D$ is the distance difference of the CD center to the reconstructed track and the true track, and $\alpha$ is the acute angle between them.

    %%%The 2D hist of Delta D and alpha
	\begin{figure*}[!htb]
        \centering
        \subfigure[The $\Delta$D distribution of single muon.]{
            \includegraphics[width=0.45\textwidth]{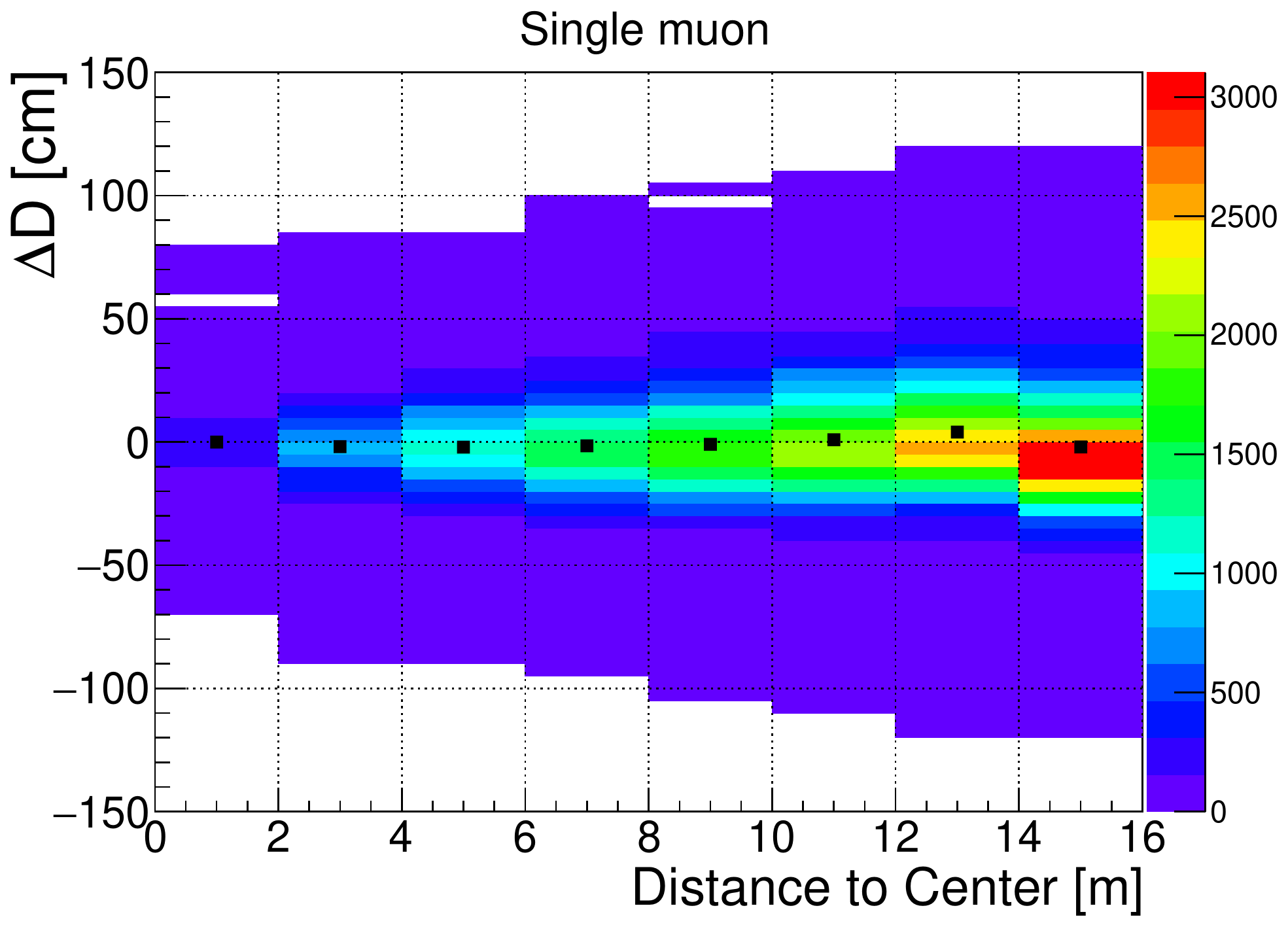}
            \label{fig: Delta D of single muon}
        }
        \quad
        \subfigure[The $\alpha$ distribution of single muon.]{
            \includegraphics[width=0.45\textwidth]{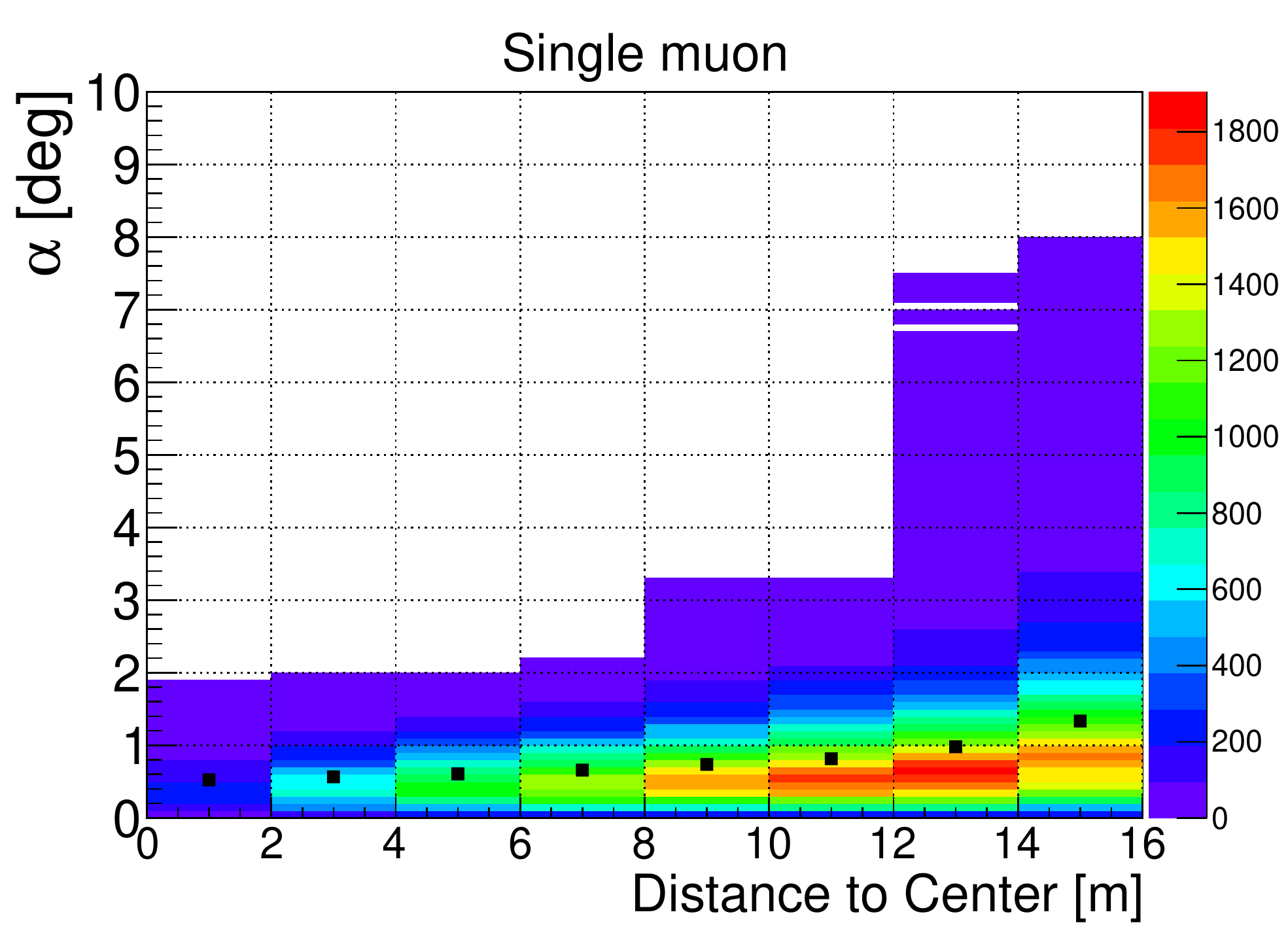}
            \label{fig: Alpha of single muon}
        }
        
        \subfigure[The $\Delta$D distribution of double muons.]{
            \includegraphics[width=0.45\textwidth]{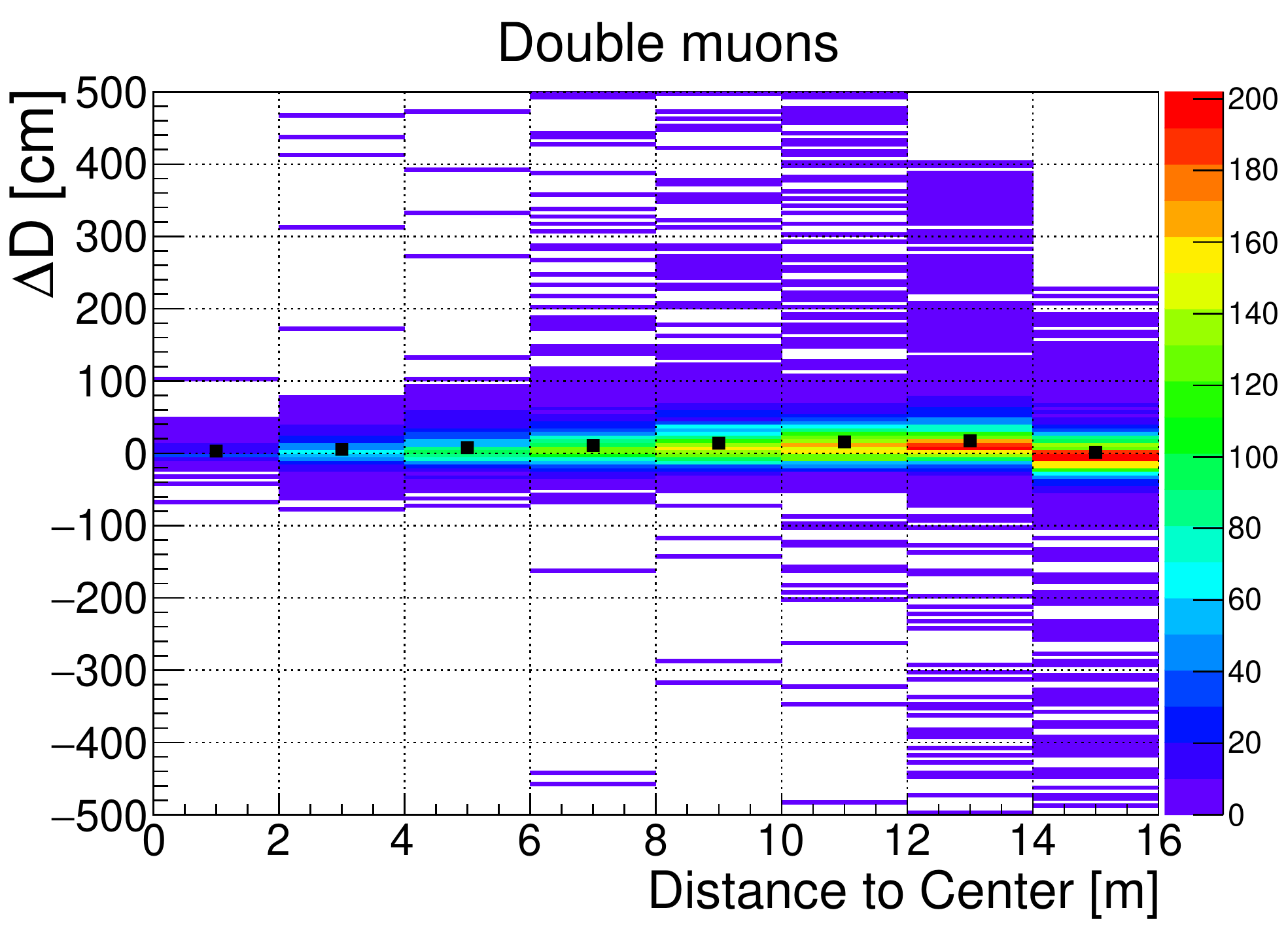}
           \label{fig: Delta D of double muon}
        }
        \quad
        \subfigure[The $\alpha$ distribution of double muons.]{
            \includegraphics[width=0.45\textwidth]{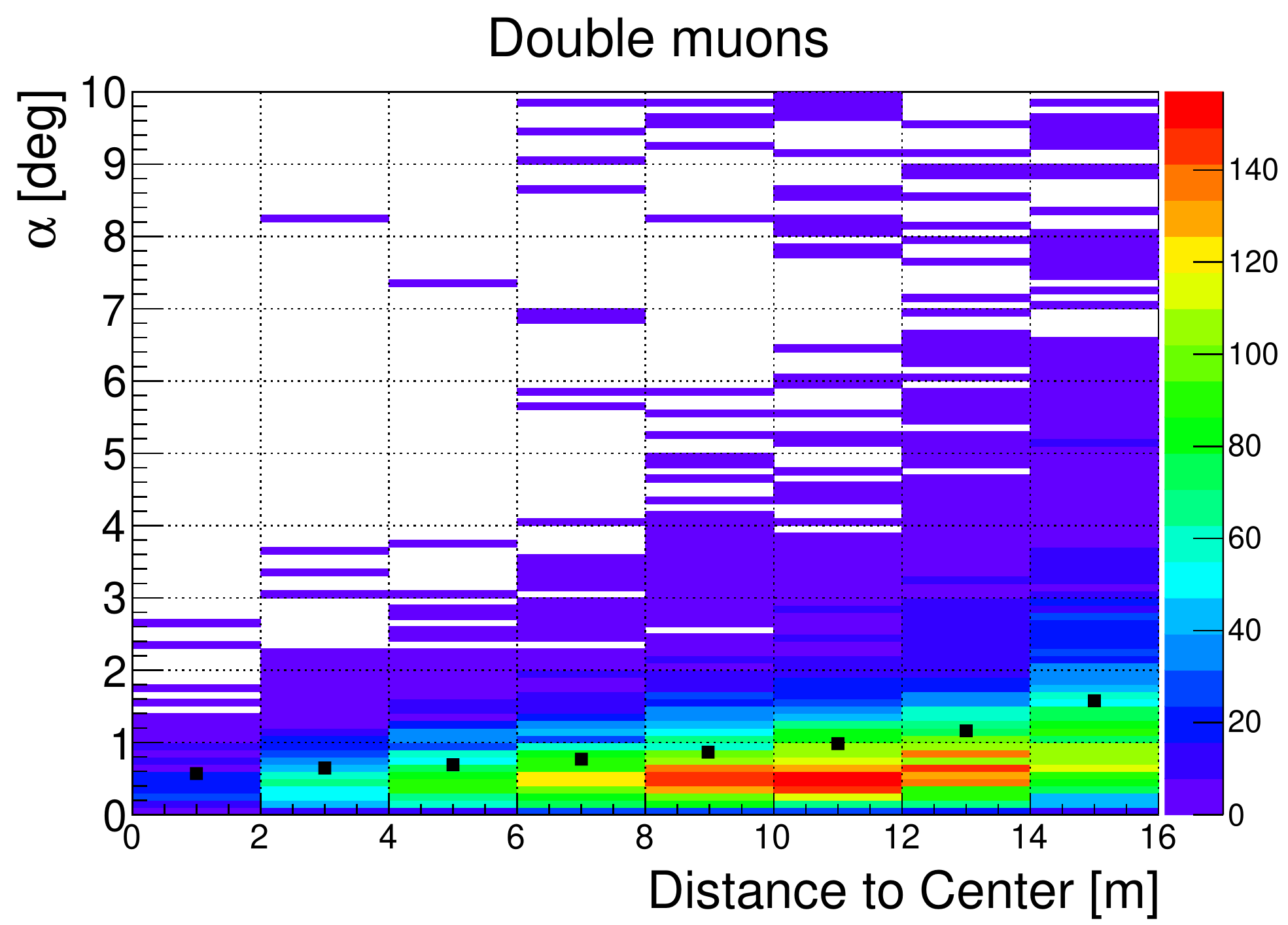}
            \label{fig: Alpha of double muon}
        }
        \caption{Reconstruction distribution of $\Delta$D and $\alpha$ versus muon distance to CD center for single muon in figure (a) and (b), for double muons in figure (c) and (d). Different distances to center indicate muon passing through different detector volumes. For double muons, $\Delta$D, $\alpha$ and the distances to the center are calculated according to their single track. In these figures, the color corresponds to the number of muon in each bin, and the black points correspond to the average values of $\Delta$D or $\alpha$ in each bin.}
        \label{fig: Reconstruction distribution of Delta D in different detector volumes}
    \end{figure*}

\begin{figure*}[!htb]
        \centering
        \subfigure[The resolution of $\Delta D$.]{
            \includegraphics[width=0.45\textwidth]{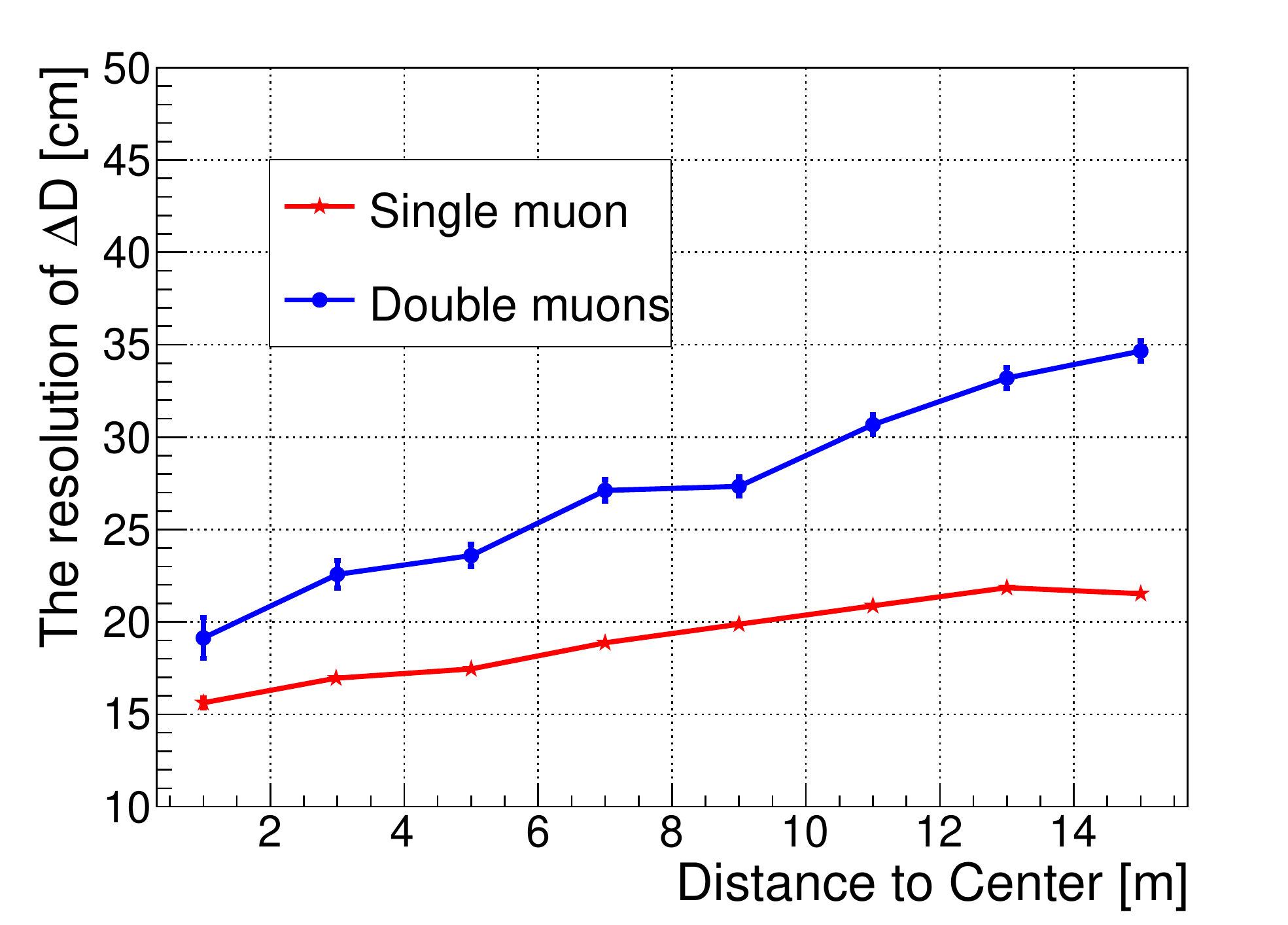}
            %\label{fig: spatial bias and resolution}
        }
        \quad
        \subfigure[The resolution of $\alpha$.]{
            \includegraphics[width=0.45\textwidth]{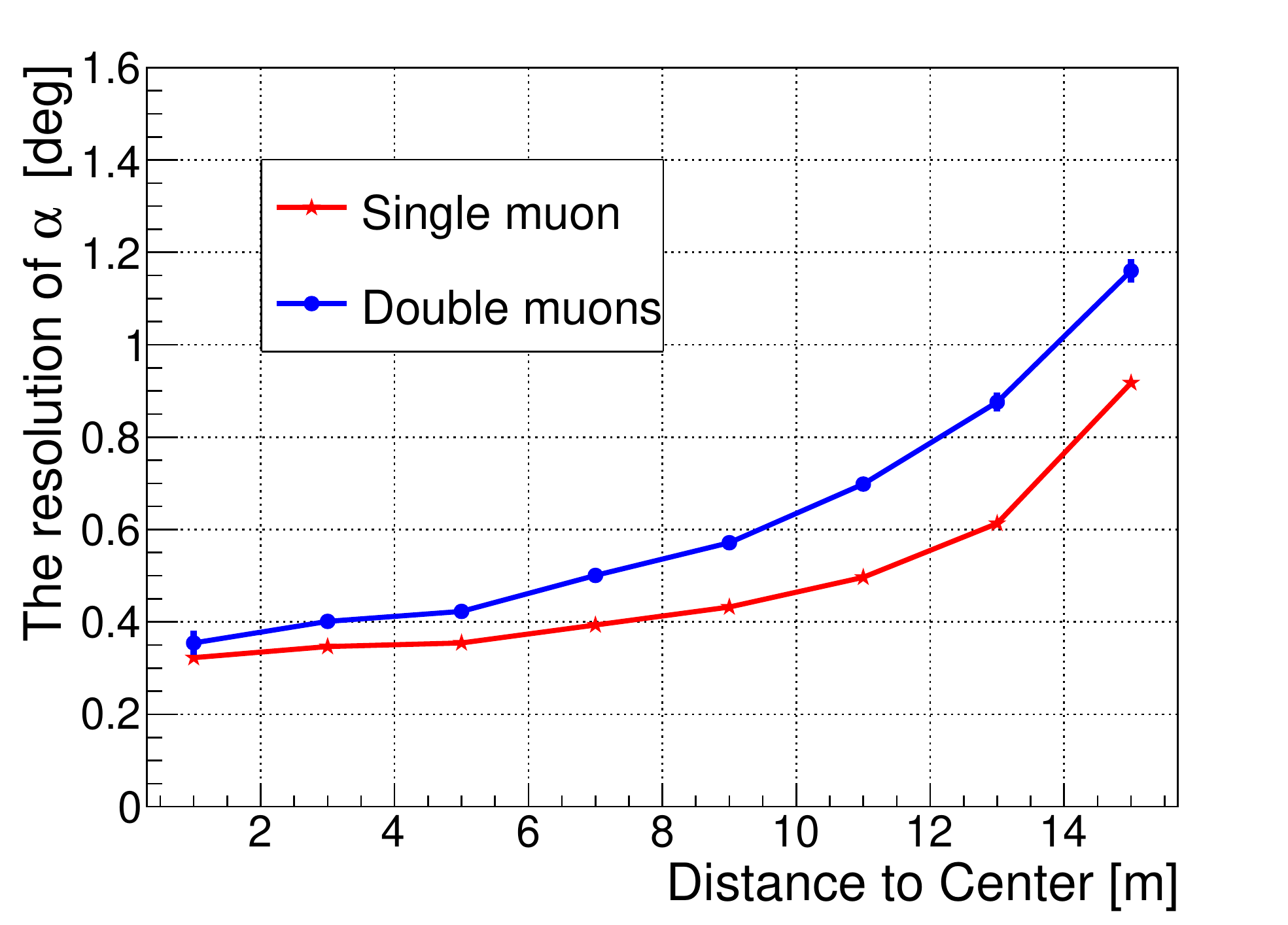}
            %\label{fig: angle bias and resolution}
        }
        \caption{The resolutions of $\Delta$D and $\alpha$ for single muon and double muons go through different detector volumes.}
        \label{fig: The resolution of Delta D and alpha}
    \end{figure*}

	Distributions of $\Delta D$ and $\alpha$ indicate the reconstruction performance. The single and double through-going muon reconstruction performance in 16~m distance to detector center is shown in Fig.~\ref{fig: Reconstruction distribution of Delta D in different detector volumes}. The distance to center is divided into 8 slices with 2~m interval, which indicate muon passing through different detector volumes. For the single muon, the average biases of $\Delta D$ in all slices (black points in Fig.~\ref{fig: Delta D of single muon}) are less than 5~cm, and the average biases of $\alpha$ (black points in Fig.~\ref{fig: Alpha of single muon}) increase from $0.5^\circ$ to $1.3^\circ$ from the detector center to the edge. As for the reconstruction of double muons, which is shown in Fig.~\ref{fig: Delta D of double muon} and Fig.~\ref{fig: Alpha of double muon}, the average biases of $\Delta D$ and $\alpha$ increase from 2~cm and $0.6^\circ$ to 15~cm and $1.5^\circ$, respectively. The reconstruction resolution of each slice is shown in Fig.~\ref{fig: The resolution of Delta D and alpha}. In the internal volume (R $\leq$ 10~m), the resolutions of $\Delta D$ and $\alpha$ are better than 20~cm and $0.5^\circ$ for single muon, and better than 30~cm and $1.0^\circ$ for double muons. In the case of muon goes through the external volume  (10~m $\leq$ R $<$ 16~m), the resolution becomes worse. This is because the track length is short near the edge and the corresponding clusters of muon incidence and exit points may have some overlap, thus there will be more deviations in the searching of cluster seeds based on the charge-weight method. As mentioned in section~\ref{section:charge response of PMT}, clipping muon (16~m $\leq$ R $<$ 17.7~m) creates short track and the two clusters are easily overlapped into one cluster, therefore they are hard to be reconstructed into one track so they can not be evaluated by the parameters $\Delta D$ and $\alpha$.
	
	%%%The resolution of Delta D and alpha
	
\subsection{Reconstruction efficiency and muon bundle veto strategy discussion}
\label{subsection:veto efficiency of reconstructed muon track}
	
	As mentioned in section~\ref{section:Reconstruction algorithm}, in order to match two candidate seeds to a muon track, several optimization methods were applied, especially in the case of more than two cluster seeds.  As a result, most of the muon tracks can be well reconstructed. However, there are still some cases where the muon track cannot be reconstructed (reconstructed track unavailable) or the reconstructed track is far away from the true track. After reconstruction, those simulated through-going muons which are without reconstructed track or whose $\Delta D$ or $\alpha$ are larger the 5~times standard deviations of their distributions in each detector volume will be tagged as failed reconstruction. Accordingly, the reconstruction efficiency can be defined as the ratio of successful reconstruction over the total events.
	
	%%%The track eff base on the Delta D and alpha
	\begin{figure*}[!htb]
		\centering
		\subfigure[]{
		    \includegraphics[width=0.45\textwidth]{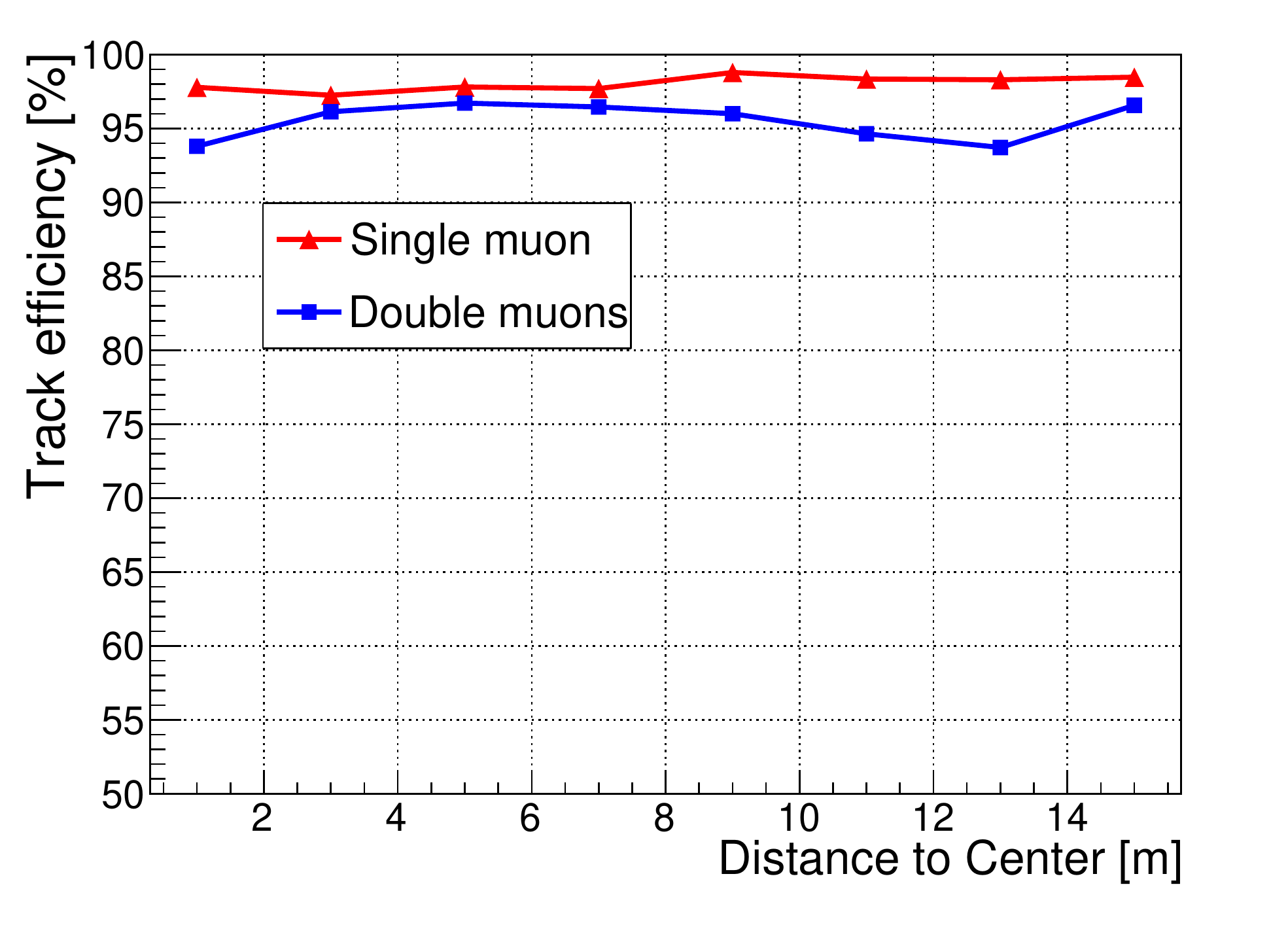}
		    \label{fig: Reconstruction efficiency_deltaD}
		}
		\quad
		\subfigure[]{
		    \includegraphics[width=0.45\textwidth]{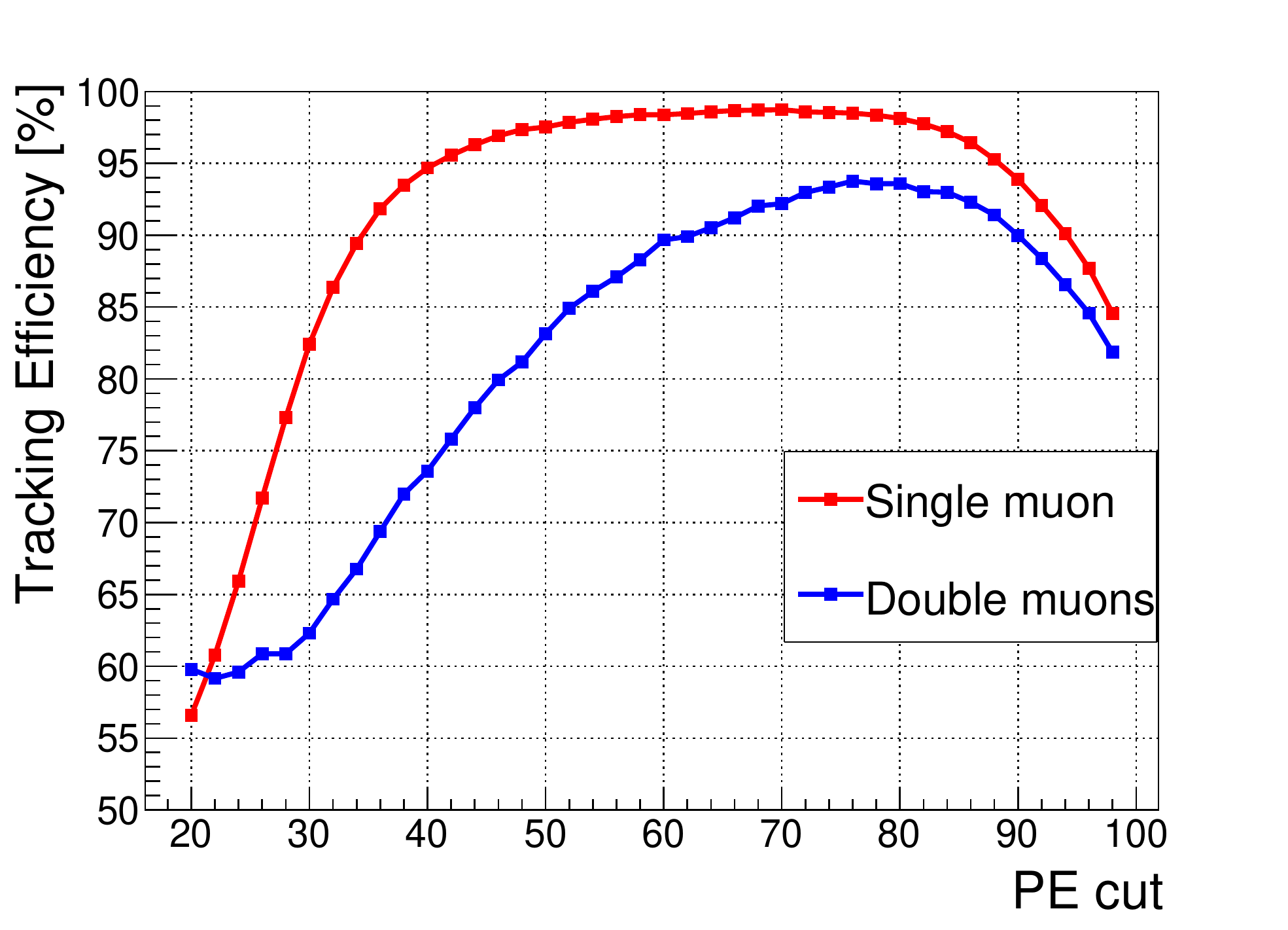}
		    \label{fig:PECUT}
		}
		\caption{(a) Reconstruction efficiency for single through-going muon and double through-going muons. For double muons, the distances to the center are calculated according to their single track. (b) The effect of PE cut (section~\ref{subsection: Charge weight method}) on reconstruction efficiency. }
	\end{figure*}
	
	Fig.~\ref{fig: Reconstruction efficiency_deltaD} shows the reconstruction efficiency of single through-going muon and double through-going muons. Single through-going muon with high reconstruction efficiency and it is around 97\% to 99\% for single muon passing through different detector volumes. In total, there are about 1.6\% of single through-going muon without reconstructed track, and about 1\% with poor reconstruction (track reconstructed but $>$ 5$\sigma$). For double through-going muons, the reconstruction efficiency is around 94\% to 97\%. And the proportions of double through-going muons tagged without track and poor reconstruction are 1.7\% and 3\%, respectively. In addition, as shown in Fig.~\ref{fig:PECUT}, the reconstruction efficiencies of single through-going muon and double through-going muons are stable when the PE cut varied from 70 PEs to 80 PEs.
	
	Muon veto strategy can be developed based on the reconstructed muon track. A simple approach is to veto a cylindrical volume along the muon trajectory in a time period~\cite{juno-yellowbook}. Considering the position dependence for reconstruction bias and resolution (section~\ref{subsection:Reconstruction performances of single and double muons}), the radius of the cylinder can be modified with the increasing of reconstruction bias and resolution~\cite{Muon_reconstruction_with_a_geometrical_model_in_JUNO}. Furthermore, a distance-dependent veto strategy was introduced in the detection of $^{8}$B solar neutrinos at JUNO~\cite{JUNO-B8}, this strategy investigates the time and distance distributions between the isotopes and muon tracks in details, and it can significantly improve the signal to background ratio. Considering the non-negligible contribution from the muon bundle and the higher estimated muon rate caused by the shallower overburden in JUNO, the veto strategy of JUNO needs to be optimized to obtain more exposure.
	
	Using the algorithm in this paper, we can reconstruct different types of muons and then develop their corresponding veto strategies in the future. For a muon event whose reconstructed track is available, a distance-dependent cylindrical veto cut along the reconstructed track will be applied. The rejected detector volume and veto time can be determined according to the space and time distributions from muon track to $^9$Li/$^8$He~\cite{JUNO-B8}. If a double muon event is reconstructed with 2 nearby reconstructed tracks (track distance d~$<$~3~m), a larger cylindrical radius will be used in the above strategy. For clipping muon or stopping muon, their tracks are difficult to reconstruct because of the hefty smearing of clusters, thus a spherical veto using the cluster seed as center will be used for conservative exclusion. As for multiple muon events, their cluster features of the charge pattern are more complicated and it is difficult to match them to the correct muon tracks. As a result, the tracking precision of multiple muons was much lower than single and double muons, so it was not shown in this paper. In reconstruction, if a muon event was tagged as multiple muons, a whole volume veto will be applied. In addition, for those failed reconstruction events whose track or cluster seed is unavailable, we will also veto a whole volume. On the other hand, a new method (named neutron veto) that can use the corresponding neutron to precisely point out the position of muon-induced isotopes is under development. Neutron veto use the sphere veto on neutron due to isotope and neutron space correlation. And it is expected to save more dead volume and allow a longer veto time window for long-lived isotope rejection. On the other hand, neutron veto is also expected to have a good ability to reject the $^9$Li/$^8$He from shower muon. Thus, besides the above muon veto, the neutron veto can be used to extend the veto strategy in the future.
	
\subsection{Reconstruction time analyze}

    \begin{figure}[ht]
		\centering
		\includegraphics[width=0.45\textwidth]{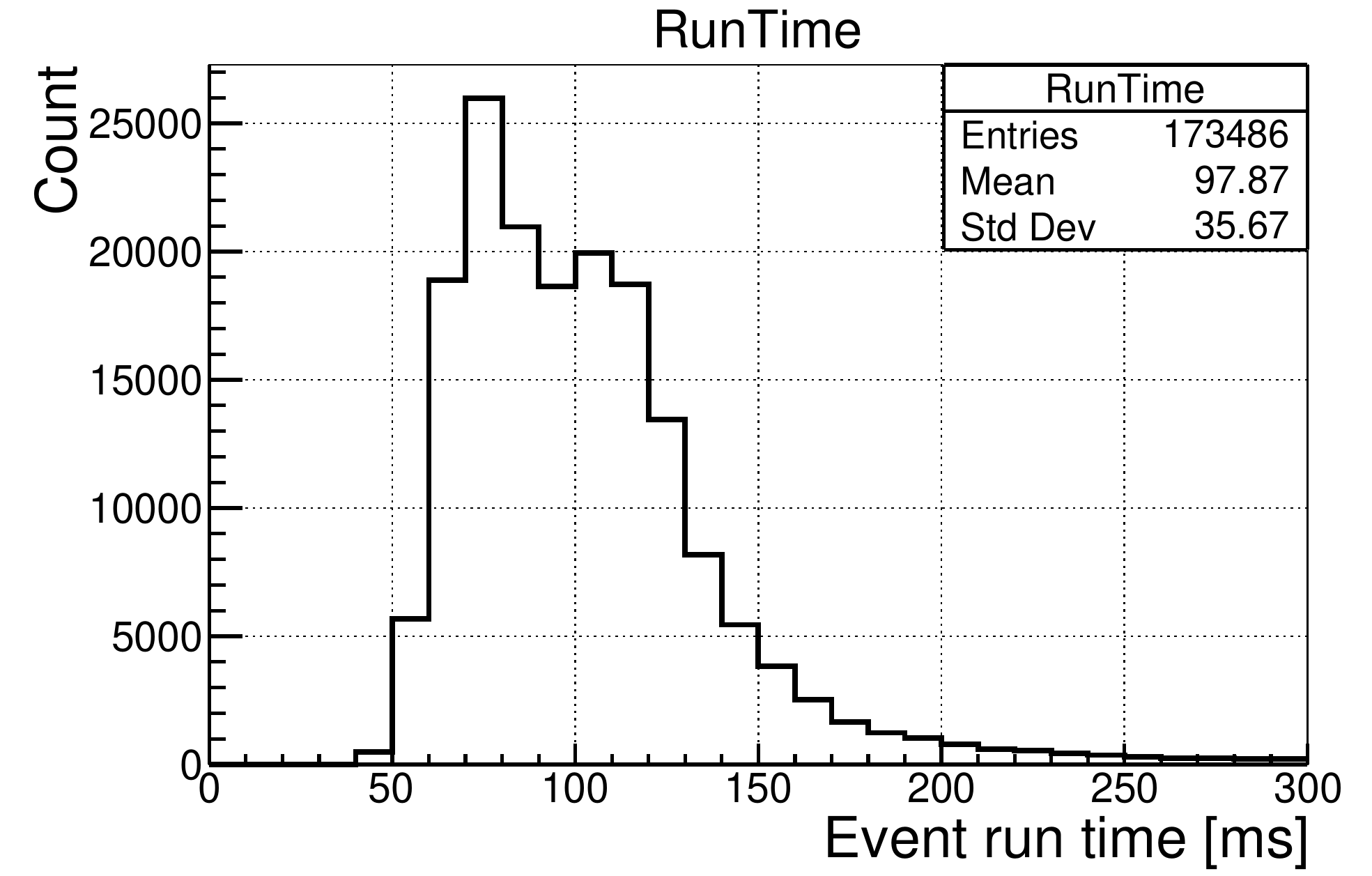}
		\caption{\label{fig: Running time} The reconstruction speed of muon event. }
	\end{figure}
	
	The time distribution of muon reconstruction is presented in Fig.~\ref{fig: Running time}, and it is about 98~ms/event. The CPU implementation of the reconstruction method in this paper provides a huge processing speed improvement of a factor of 51 compared to the fastest light method (5000~ms/event)~\cite{the_fastest_light_in_the_JUNO_central_detector,a_convolutional_neural_network}. This speed makes it possible to apply the algorithm in the online event classification, which requires a fast online-filtering for a possible data rate reduction. 
	
\section{Summary}
	
	Muon bundles are rarely reconstructed in the previous underground low event rate experiments. We developed a muon bundle reconstruction algorithm, mainly focusing on single and double muons, based on JUNO underground muon samples simulated by Geant4 in the JUNO offline software. For the spherical geometry of JUNO's LS target, different muons have different characteristics. Muon events are classified into the single, double, multiple muon categories and each has through-going, clipping and stopping features. These kinds of muons are considered and reconstructed, which is more like the real situation in the future when taking the data.   
	
	The muon track incident and exit clusters on the PMT charge hit pattern can be roughly searched and their centers can be fitted using the  ROOT tool `TSpectrum2'. However, the so-obtained clusters have to be further processed and corrected. Besides re-do charge weighting based on surround PMTs, several optimization methods, such as rotation projection, charge smoothing, geometry correction were developed. The complicated clustering strategy of the double muons is discussed in detail. The reconstructed results have indicated that the algorithm can reconstruct muon tracks with a resolution of 20~cm in distance to detector center, $0.5^\circ$ in angle, and that the tracking efficiency is 98\% for single muon events. For double muon events, the resolutions of distance to detector center and angle are about 30~cm and $1.0^\circ$, respectively, and the tracking efficiency is 95\%. This is the first reconstruction of muon bundles in a large volume liquid scintillator detector and the algorithm in this paper shows good performance and potential.
	
	Finally, we discussed an optimized veto strategy that includes distance-dependent cylindrical veto and spherical veto for different muons, and it is expected to obtain more exposure. The reconstruction speed is fast compared to reconstruction methods that use timing information. This advantage can promote the algorithm to be deployed in the online event classification which needs fast calculation speed to classify events in order to perform a fast online-filtering for a possible data rate reduction.

\section{Acknowledgments}
	We thank the JUNO offline and reconstruction working group for many helpful discussions. This work was supported by National Natural Science Foundation of China No. 12005044 and 11975258, the Strategic Priority Research Program of the Chinese Academy of Sciences, Grant No. XDA10011200 and XDA10010900.

\iffalse	
\begin{flushleft}
	{\bf Author Contributions} All authors contributed to the study's conception and design. Material preparation, data collection, and analysis were performed by Cheng-Feng Yang, Yong-Bo Huang, Ji-Lei Xu, Di-Ru Wu, Hao-Qi Lu, Yong-Peng Zhang, Wu-Ming Luo, Miao He, Guo-Ming Chen, and Si-Yuan Zhang. The first draft of the manuscript was written by Cheng-Feng Yang, Yong-Bo Huang, and Ji-Lei Xu, and all authors commented on previous versions of the manuscript. All authors read and approved the final manuscript.
\end{flushleft}
\fi
	
% ----------------------------------------------------------------

	% ----------------------------------------------------------------
\end{document}